\documentclass[11pt,letterpaper]{article}
\usepackage[margin=0.85in,includeheadfoot]{geometry}
\usepackage[T1]{fontenc}
\usepackage{lmodern} 
\usepackage[utf8]{inputenc}
\usepackage[round,authoryear]{natbib}
\usepackage{microtype}
\usepackage{caption}

\usepackage{amsmath,amsfonts,amssymb}
\usepackage{graphicx}
\usepackage{float} 
\usepackage[american]{babel}
\usepackage{booktabs}
\usepackage[section]{placeins}
\usepackage{enumitem}
\usepackage{mathtools}
\usepackage{bm}
\usepackage{dsfont}
\usepackage{bbm}
\usepackage{xcolor}
\usepackage{url}
\usepackage[colorlinks,citecolor=blue,urlcolor=blue,linkcolor=blue]{hyperref}
\hypersetup{pdftitle={A Subjective Bayesian Approach to Address Unchecked Assumptions in a Causal Analysis},pdfauthor={Jaiyool Kim, Indrabati Bhattacharya, Jonathan R. Bradley},pdfkeywords={Bayesian statistics, Causal inference, Mixture models, Unchecked assumptions, Unmeasured confounding}}

\definecolor{revisiongreen}{RGB}{0,120,0}
\definecolor{numericalrevisionteal}{RGB}{0,100,100}
\definecolor{pinegreen}{RGB}{0,120,100}

\providecommand{\suppgraphic}[2][0.95\textwidth]{%
  \includegraphics[width=#1]{#2}%
}

\newcommand{\vect}[1]{\bm{#1}}
\newcommand{\mat}[1]{\mathbf{#1}}

\title{A Subjective Bayesian Approach to Address Unchecked Assumptions in a Causal Analysis}
\author{Jaiyool Kim$^{1,*}$, Indrabati Bhattacharya$^{1,**}$,\\
Jonathan R. Bradley$^{2,***}$\\[0.6em]
{\normalsize $^1$Department of Statistics, Florida State University}\\
{\normalsize $^2$Statistics and Data Science, University of Missouri}\\[0.35em]
{\small $^*$jk21y@fsu.edu \quad $^{**}$ib22g@fsu.edu \quad $^{***}$bradleyjr@missouri.edu}}
\date{}

\begin{document}
\maketitle

\begin{abstract}
Causal analyses using observational data rely on identifying assumptions that cannot be verified from the observed data alone. When such unchecked assumptions fail, effect estimates may be biased and uncertainty may be understated. We propose the Adjusted Model for Unchecked Assumptions (AMUA), a subjective Bayesian framework that propagates uncertainty about the joint validity of key causal assumptions through a two-component posterior mixture. One component corresponds to the standard observational causal model in which the assumptions hold (referred to as the ``assumption-hold component''); the other represents a flexibly defined model that allows the causal assumptions to fail (referred to as the ``assumption-fail component''). The mixture weight $\pi$ is specified by the analyst as the prior probability assigned to the assumption-hold component. This formulation leads to a new diagnostic, $\pi_{\min}$: the smallest value of $\pi$ for which the 95\% HPD set for the average treatment effect (ATE) excludes zero. In simulations with a spatially structured unmeasured confounder, we evaluate bias, interval length, and coverage as $\pi$ varies. The results show that acknowledging assumption uncertainty can widen intervals, and improve coverage relative to analyses that treat the assumptions as certain. In a county-level application relating long-term PM$_{2.5}$ exposure to COVID-19 mortality, AMUA shows how posterior conclusions depend on prior confidence in the identifying assumptions.
\end{abstract}



\section{Introduction}\label{ch2:sec:introduction}

{
A central difficulty in causal inference with observational data is that the average treatment effect (ATE) cannot be identified from the observed data alone. Within the potential-outcome framework, causal conclusions require assumptions that are substantively important yet not empirically verifiable from the data at hand \citep{Rubin1974,ImbensRubin2015,HernanRobins2020}. Here we focus on four such identifying assumptions:\begin{enumerate}
\item[\textbf{A1:}] \textbf{SUTVA.} There is no interference between units, and treatment is well defined as a single version.
\item[\textbf{A2:}] \textbf{Consistency.} The observed response equals the potential outcome corresponding to the observed treatment level.
\item[\textbf{A3:}] \textbf{Latent ignorability.} Conditional on observed covariates and random effects, the potential outcomes are independent of the treatment assignment.
\item[\textbf{A4:}] \textbf{Positivity.} Each treatment level has positive probability on the relevant covariate support.
\end{enumerate} 
Since these assumptions are not testable from observational data alone, observational causal analyses remain inherently vulnerable to violation of these assumptions \citep{ImbensRubin2015,HernanRobins2020,ReichYangGuan2021}. To address this uncertainty, we propose the Adjusted Model for Unchecked Assumptions (AMUA), a Bayesian framework that places prior probability on the joint validity of A1--A4 and propagates this assumption uncertainty through posterior inference.

We are motivated by a reanalysis of a consequential public-health application involving the widely studied association between long-term PM${2.5}$ exposure and COVID-19 mortality, which has been interpreted causally. An influential U.S. county-level ecological regression analysis reported a positive association between long-term PM${2.5}$ exposure and COVID-19 mortality \citep{WuNetherySabathBraunDominici2020}, and spatial causal inference methods have subsequently been used to revisit this environmental-health question \citep{ReichYangGuan2021}. We ask a different question: how much prior confidence in the joint validity of A1--A4 is required for the PM$_{2.5}$--COVID-19 causal conclusion to remain? We also investigate whether a spatial random-effects model can be used to represent prior beliefs about the validity of A1--A4 in this setting. AMUA thus provides not only a baseline causal conclusion from a posterior distribution, but also a diagnostic of how strongly that conclusion depends on prior confidence in the identifying assumptions.

A major response to uncertainty about unchecked causal assumptions is sensitivity analysis. Sensitivity analysis assesses how conclusions depend on causal assumptions that cannot be verified from the observed data alone. The AMUA is partially motivated by Tukey’s factorization, a reweighting framework that separates the observed-data distribution from an unidentified selection mechanism \citep{FranksDAmourFeller2020}. Tukey’s factorization introduces sensitivity parameters that impose additional assumptions on the relationship between treatment assignment and potential outcomes, thereby characterizing the magnitude of unmeasured confounding. Related approaches quantify sensitivity to unmeasured confounding through parameters describing departures from identification or the strength of confounder associations \citep{RichardsonEvansRobins2010}. Other methods provide more specialized diagnostics. For example, the E-value quantifies the minimum strength of association that an unmeasured confounder would need to have with treatment and outcome to explain away an observed association \citep{VanderWeeleDing2017}, while the Impact Threshold for a Confounding Variable (ITCV) and Robustness of Inference to Replacement (RIR) assess the magnitude of unmeasured confounding or bias required to overturn an inference. Other methods seek to address specific sources of bias by imposing alternative assumptions or additional structures. For example, negative-control methods can be used to detect and, under additional assumptions, address confounding bias. Proxy-variable methods can identify causal effects when suitable proxy variables satisfy conditions such as proxy conditional independence and rank or completeness conditions \citep{LipsitchTchetgenCohen2010,Tchetgen2014COCA,MiaoGengTchetgen2018}. Some consider introducing an Instrumental-variable (IV) to adjust the model for identifying assumptions. IV methods seek to identify causal effects using variation in treatment induced by an instrument. The instrument must be related to treatment and affect the outcome only through treatment \citep[see, e.g.,][for the conditions]{FuKim2021,BaiocchiChengSmall2014}. These approaches address particular sources of bias or rely on specific identifying structures rather than directly representing uncertainty about the joint validity of multiple causal assumptions.

AMUA complements these approaches by explicitly representing uncertainty about the joint validity of A1–A4 through a Bayesian mixture model. This motivates another use of the AMUA, where one finds the value of the prior probability of A1--A4 for which AMUA is ``close'' to an existing statistical model's posterior distribution. That is, one can calibrate the prior belief of A1--A4 in the AMUA so that its posterior distribution is ``close'' (in integrated squared error) to an existing model that addresses at least one of A1--A4. This calibrated value of the mixture weight acts as a diagnostic that aids with interpreting how a causal model implicitly places a prior on A1--A4. We explore this use of the AMUA in our empirical results.

AMUA represents uncertainty about whether A1--A4 jointly hold through a two-component mixture model. The first component corresponds to the standard observational causal analysis, in which A1--A4 are treated as valid; the second component is a flexible assumption-fail component, in which one or more of these assumptions may be unreliable. Unlike sensitivity analyses that parameterize specific departures from identification, AMUA accommodates uncertainty about assumption failure without specifying a single failure mechanism. As detailed in Section~\ref{ch2:sec:models}, its formal construction preserves the intended component marginals while ensuring that the observed-data likelihood does not distinguish between the assumption-hold and assumption-fail components. This allows uncertainty about the joint validity of the identifying assumptions to propagate into posterior inference for the causal estimand. Thus, AMUA is not intended to necessarily improve estimation of the ATE, but rather to assess how causal conclusions depend on uncertainty about the validity of the identifying assumptions.

AMUA's formulation is intentionally conservative. As less prior weight is placed on the assumption-hold component, posterior intervals may widen, and posterior evidence for a nonzero ATE may weaken. We view this as a feature rather than a defect when the identifying assumptions are genuinely uncertain: incorporating this uncertainty can reduce overconfidence that may arise when the assumptions are treated as known to hold. The formulation also leads to a new inferential question: what is the smallest prior belief in the joint validity of A1--A4 for which the posterior summary for the ATE still excludes zero? We summarize this threshold by $\pi_{\min}$, which links prior belief in the identifying assumptions to posterior evidence about the causal effect.

There are three main contributions of this paper. First, AMUA provides a Bayesian mechanism for propagating uncertainty about unchecked identifying assumptions directly into posterior inference, rather than conditioning on A1--A4 as known to hold. Second, we introduce $\pi_{\min}$, a diagnostic summary that quantifies the minimum prior probability assigned to the joint validity of A1--A4 required for the ATE's Highest Posterior Density (HPD) interval to exclude zero under a chosen HPD criterion. We define $\pi_{\min}$ formally in Section~2.6. Third, the prior probability assigned to A1--A4 can be calibrated so that AMUA approaches an existing model, such as a spatial random-effects model, providing a diagnostic of the prior belief in A1--A4 implicitly encoded by that model. The PM$_{2.5}$--COVID-19 mortality application illustrates these ideas in a substantive observational setting.

The paper is organized as follows. Section~\ref{ch2:sec:models} introduces the AMUA framework and its two-component mixture representation. Section~\ref{ch2:sec:simulation} presents the simulation study. Section~\ref{ch2:sec:application} presents the PM$_{2.5}$--COVID-19 application, and Section~\ref{ch2:sec:discussion} concludes.

}


\section{Models}
\label{ch2:sec:models}

\subsection{Preliminaries}

We first define the notation for the observed data and latent quantities in AMUA. For unit $i=1,\ldots,n$, let $A_i\in\{0,1\}$ denote the observed treatment, let $Z_i\in\mathcal Z$ denote the observed outcome, $\mathbf{x}_i\in\mathbb{R}^{p}$ the covariates, and $\kappa$ model-specific latent random effects, such as spatial effects, defined more fully below. Let
$\mathbf z=(Z_1,\ldots,Z_n)^\top$,
$\mathbf a=(A_1,\ldots,A_n)^\top$, and
$\mathbf X=(\mathbf x_1,\ldots,\mathbf x_n)^\top$ denote the corresponding data vectors. We use ``$\sim$'' to distinguish quantities in the assumption-fail component from their counterparts in the standard observational model. For example, let
$\tilde{\mathbf a}=(\tilde A_1,\ldots,\tilde A_n)^\top\in\{0,1\}^n$
denote the treatment-assignment vector in the assumption-fail component, and let
$\tilde{\mathbf a}_{-i}$ denote the assignments for units other than $i$.
For $a\in\{0,1\}$, define
$
\tilde Z_i(a,\tilde{\mathbf a}_{-i})\in\mathcal Z$
as the potential outcome for unit $i$ when its treatment level is
set to $a$. We allow the potential outcome $\tilde{Z}$ to be affected by the assignments of other units as it is a quantity in our assumption fail component of our mixture model. For $j\in\{0,1\}$, define
$ 
\tilde{\mathbf z}_{j}
=
\big(
\tilde Z_1(j,\tilde{\mathbf a}_{-1}),\ldots,
\tilde Z_n(j,\tilde{\mathbf a}_{-n})
\big)^\top.
$
Thus, $\tilde{\mathbf z}_{j}$ collects the latent potential outcomes obtained
by setting every focal unit's own treatment argument to the common level $j$,
while retaining the possibly relevant treatment configuration of the other
units.
We also define 
$
\tilde{\mathbf z}
=
\big(
\tilde Z_1(\tilde A_1,\tilde{\mathbf a}_{-1}),
\ldots,
\tilde Z_n(\tilde A_n,\tilde{\mathbf a}_{-n})
\big)^\top.
$
Thus, $\tilde{\mathbf z}$ collects the latent potential
outcomes evaluated under $\tilde{\mathbf a}$. 
The target causal estimand is the average treatment effect
\begin{align}
\mathrm{ATE}
:=
\mathbb{E}_{\mathbf X, \tilde{\mathbf A}, \tilde{\mathbf Z}}
\left\{
\tilde{Z}_i(1,\tilde{\mathbf a}_{-i})
-
\tilde{Z}_i(0,\tilde{\mathbf a}_{-i})
\right\}.
\label{ch2:eq:ate_main} 
\end{align}

The first arguments $1$ and $0$ denote the two well-defined own-treatment
levels, rather than values of a separate latent treatment-version variable. Equation~\eqref{ch2:eq:ate_main} is the population-level
target that AMUA summarizes under uncertainty about the causal assumptions.

The standard observational causal analysis relies on four identifying assumptions, labeled A1--A4 (defined in \ref{ch2:sec:introduction}). Together, A1--A4 connect the observed-data model to the target ATE in \eqref{ch2:eq:ate_main} within the assumption-hold component. To represent uncertainty about their joint validity, we introduce a latent component indicator $
\delta \sim \mathrm{Bernoulli}(\pi)$, where $\pi=\Pr(\delta=1)$ is the analyst's prior probability, or mixture weight, assigned to the assumption-hold component. When $\delta=1$, inference proceeds under the standard observational causal model satisfying A1--A4; when $\delta=0$, inference proceeds under a more flexible model in which one or more assumptions may fail. We refer to this two-component formulation as the Adjusted Model for Unchecked Assumptions (AMUA). The indicator $\delta$ is not identified by the observed-data likelihood. It represents the analyst's prior uncertainty about the joint validity of the identifying assumptions, propagating this uncertainty into posterior inference for the causal effect. The target ATE in \eqref{ch2:eq:ate_main} is not conditioned on $\delta$ but instead marginalizes over $\delta$. Thus, $\delta$ indexes uncertainty about the identifying assumptions rather than defining a separate causal estimand for each component. Although $\delta$ is latent, Component 1 is formulated directly for the observed data, whereas Component 2 introduces auxiliary latent quantities to represent possible assumption-failure configurations.

A useful feature of this setup is that it allows the analyst to assign prior mass to different configurations of assumption validity. In particular, $\pi$ can be interpreted as the prior probability that all four identifying assumptions hold simultaneously:
\begin{align*}
\pi = \Pr(\delta=1)=\Pr(\text{A1}\cap\text{A2}\cap\text{A3}\cap\text{A4}),
\end{align*}
where the assumptions are treated as events under the analyst's prior probability measure. More generally, the analyst may assign prior mass to different validity configurations of $(\mathrm{A1},\mathrm{A2},\mathrm{A3},\mathrm{A4})$. By the law of total probability,
\begin{align*}
1
&=
\Pr(\mathrm{A1}\cap\mathrm{A2}\cap\mathrm{A3}\cap\mathrm{A4})
+
\sum_{c\in\mathcal{C}_{\mathrm{fail}}}\Pr(c),
\end{align*}
where $\mathcal{C}_{\mathrm{fail}}$ denotes the collection of mutually exclusive configurations in which at least one of A1--A4 fails. Thus $\pi= \Pr(\text{A1}\cap\text{A2}\cap\text{A3}\cap\text{A4})$ is the prior mass assigned to the assumption-hold configuration, while $1-\pi  = \sum_{c\in\mathcal{C}_{\mathrm{fail}}} \Pr(c)$ is the total prior mass assigned to all assumption-failure configurations combined. In our formulation, any assumption-failure configuration includes failure of A2 (consistency). This allows the latent potential outcomes to be modeled separately from the observed outcomes, without imposing equality between the observed outcome and the potential outcome corresponding to the observed treatment. The framework can nevertheless accommodate multiple assumption-failure configurations. For example, an analyst might regard A3 and A4 as highly credible while remaining more doubtful about A1 and A2. One possible prior specification is
$
\Pr(\mathrm{A1}\cap\mathrm{A2}\cap\mathrm{A3}\cap\mathrm{A4})=0.70$
and
$\Pr(\mathrm{A1}^{c}\cap\mathrm{A2}^{c}\cap\mathrm{A3}\cap\mathrm{A4})=0.25$,
with the remaining 0.05 prior mass distributed over other configurations contained in $\mathrm{A2}^{c}$. Under this specification, the assumption-hold component receives mixture weight 0.70, while the assumption-fail component receives the remaining 0.30. This example illustrates how prior beliefs about individual assumptions can be represented through a distribution over validity configurations.

\subsection{Component 1: Observational model when A1--A4 hold}

Component 1 is the parametric observational model under the assumption that A1--A4 hold. The observed outcome, treatment, and random effects are modeled conditional on parameter blocks $\boldsymbol{\theta}$ and $\omega$, which govern the outcome and the treatment-assignment models, respectively. The model consists of four building blocks:
\begin{align}
\text{Observed Outcome Model:}\quad & f_{\mathrm{obs}}(\mathbf z\mid \mathbf a,\kappa,\boldsymbol{\theta},\mathbf X), \nonumber\\
\text{Observed Treatment Model:}\quad & f_{\mathrm{obs}}(\mathbf a\mid \omega,\mathbf X), \nonumber\\
\text{Random Effects Model:}\quad & f(\kappa\mid \boldsymbol{\theta},\mathbf X), \nonumber\\
\text{Parameter Model:}\quad & f(\boldsymbol{\theta}\mid \mathbf X)\,f(\omega\mid \mathbf X). \label{ch2:eq:obs_blocks}
\end{align}
The outcome and treatment models are specified to support the standard causal interpretation under A1--A4. Multiplying the four building blocks in \eqref{ch2:eq:obs_blocks} gives the Component 1 joint distribution
\begin{align}
f_{\mathrm{obs}}(\mathbf z,\kappa,\mathbf a,\boldsymbol{\theta},\omega\mid \mathbf X)
= f_{\mathrm{obs}}(\mathbf z\mid \mathbf a,\kappa,\boldsymbol{\theta},\mathbf X)
 f_{\mathrm{obs}}(\mathbf a\mid \omega,\mathbf X)
 f(\kappa\mid \boldsymbol{\theta},\mathbf X)
 f(\boldsymbol{\theta}\mid \mathbf X)
 f(\omega\mid \mathbf X).
\label{ch2:eq:obs_joint_generic}
\end{align}
In what follows, $f_{\mathrm{obs}}$ refers to either the observational building blocks in \eqref{ch2:eq:obs_blocks} or their joint distribution in \eqref{ch2:eq:obs_joint_generic}. The detailed hierarchical specification is given in Section~\ref{ch2:subsec:app-models}.

\subsection{Component 2: Assumption-fail component and component-specific ATE}

Component 2 is designed to remain flexible when one or more of A1--A4 may fail. In this component, the latent potential outcomes may depend on the full treatment configuration, and the treatment versions need not coincide with the observed $A_i$. Conditional on the component indicator $\delta$, the latent building blocks are
\begin{align}
\text{Potential Outcome Model:}\quad
&\prod_{j=0}^{1} f(\tilde{\mathbf z}_j\mid \tilde{\mathbf a},\tilde{\kappa},\widetilde{\boldsymbol{\theta}},\mathbf X,\delta), \nonumber\\
\text{Latent Outcome Model:}\quad
& f(\tilde{\mathbf z}\mid \tilde{\mathbf a},\tilde{\kappa},\widetilde{\boldsymbol{\theta}},\tilde{\omega},\mathbf X,\delta), \nonumber\\
\text{Latent Treatment Model:}\quad
& f(\tilde{\mathbf a}\mid \tilde{\kappa},\widetilde{\boldsymbol{\theta}},\tilde{\omega},\mathbf X,\delta), \nonumber\\
\text{Random Effects Model:}\quad
& f(\tilde{\kappa}\mid \widetilde{\boldsymbol{\theta}},\tilde{\omega},\mathbf X,\delta), \nonumber\\
\text{Parameter Model:}\quad
& f(\tilde{\omega}\mid \widetilde{\boldsymbol{\theta}},\mathbf X,\delta)\, f(\widetilde{\boldsymbol{\theta}}\mid \mathbf X,\delta).
\label{ch2:eq:latent_blocks}
\end{align}

The factorization does not impose conditional independence between treatment and potential outcomes. Dependence is permitted through the conditioning of the potential-outcome model on $\tilde{\mathbf a}$ and through the shared latent quantities $\tilde{\kappa}$ and $\widetilde{\boldsymbol{\theta}}$. The resulting joint distribution for Component 2 is
\begin{align*}
&f_{\mathrm{mix}}(\tilde{\mathbf z}_0,\tilde{\mathbf z}_1,\tilde{\mathbf z},
\tilde{\kappa},\widetilde{\boldsymbol{\theta}},\tilde{\omega},\tilde{\mathbf a},
\delta=0\mid \mathbf X)\\
&=\prod_{j=0}^{1} f(\tilde{\mathbf z}_j\mid \tilde{\mathbf a},\tilde{\kappa},\widetilde{\boldsymbol{\theta}},\mathbf X,\delta=0)f(\tilde{\mathbf z}\mid \tilde{\mathbf a},\tilde{\kappa},\widetilde{\boldsymbol{\theta}},\tilde{\omega},\mathbf X,\delta=0)f(\tilde{\mathbf a}\mid \tilde{\kappa},\widetilde{\boldsymbol{\theta}},\tilde{\omega},\mathbf X,\delta=0)\\
&\hspace{20pt}\times f(\tilde{\kappa}\mid \widetilde{\boldsymbol{\theta}},\tilde{\omega},\mathbf X,\delta=0)f(\tilde{\omega}\mid \widetilde{\boldsymbol{\theta}},\mathbf X,\delta=0)\, f(\widetilde{\boldsymbol{\theta}}\mid \mathbf X,\delta=0)(1-\pi).
\end{align*}

\noindent This component is not intended to encode one specific failure mechanism in
advance. Its joint law can accommodate interference through dependence of 
$\tilde Z_i(j,\tilde{\mathbf a}_{-i})$, $j\in\{0,1\}$, on other units' assignments; departures from latent ignorability can occur through dependence
between $(\tilde{\mathbf z}_0,\tilde{\mathbf z}_1)$ and
$\tilde{\mathbf a}$ in the joint factorization; and positivity violations can occur through a treatment law that assigns
little or no probability to some treatment--covariate combinations. The model in \eqref{ch2:eq:latent_blocks} is therefore intentionally left largely unspecified, allowing the framework to accommodate a broad class of assumption-failure mechanisms.

Since the assumption-fail component is intentionally broad, we impose only minimal structure needed to keep the target estimand interpretable across components. Specifically, we include treatment-specific coefficient vectors $\widetilde{\boldsymbol\beta}^{(1)}$ and $\widetilde{\boldsymbol\beta}^{(0)}$ in the latent parameter block $\widetilde{\boldsymbol\theta}$. We directly specify only the prior distributions of $\widetilde{\boldsymbol\beta}^{(1)}$ and $\widetilde{\boldsymbol\beta}^{(0)}$, and the remaining latent distributions are left sufficiently flexible to allow for different assumption-failure configurations.

We therefore define a component-specific ATE, denoted by $\mathrm{ATE}_{\mathrm{CS}}$. The word component refers to the mixture component indexed by $\delta=1$ or $\delta=0$. We define
\begin{align}
\mathrm{ATE}_{\mathrm{CS}}\big(\widetilde{\boldsymbol\beta}^{(1)},\widetilde{\boldsymbol\beta}^{(0)},\delta\big)
:={}&
\mathbb{E}_{\mathbf X,\tilde{\mathbf A}, \tilde{\mathbf Z}}
\left[
\tilde{Z}_i(1,\tilde{\mathbf a}_{-i})-
\tilde{Z}_i(0,\tilde{\mathbf a}_{-i})
\mid
\widetilde{\boldsymbol\beta}^{(1)},\widetilde{\boldsymbol\beta}^{(0)},\delta
\right].
\label{ch2:eq:atecs_delta_def}
\end{align}
This quantity differs from the main ATE in \eqref{ch2:eq:ate_main}, which is not conditioned on $\delta$ or the latent coefficient values. Thus, AMUA retains a common causal target while allowing uncertainty about the identifying assumptions to enter its posterior distribution through the mixture components.

The population ATE in \eqref{ch2:eq:ate_main} remains the main counterfactual of the paper. However, we need $\mathrm{ATE}_{\mathrm{CS}}$ in our new framework, since when combining the two components by marginalizing across $\delta$, each component must map its
parameters to the same form of the component specific ATE. For the
assumption-fail component, equation~\eqref{ch2:eq:atecs0_beta_relation} supplies
this common mapping by expressing the average treatment contrast as the
population-average difference between the two treatment-specific regression
surfaces. That is, we require the component-specific ATE in \eqref{ch2:eq:atecs_delta_def} to satisfy
\begin{align}
\mathrm{ATE}_{\mathrm{CS}}\big(\widetilde{\boldsymbol\beta}^{(1)},\widetilde{\boldsymbol\beta}^{(0)},\delta\big)
=
\mathbb{E}(\mathbf x^\top\vert \delta)\big(\widetilde{\boldsymbol\beta}^{(1)}-\widetilde{\boldsymbol\beta}^{(0)}\big).
\label{ch2:eq:atecs0_beta_relation}
\end{align}
This requirement specifies only the common scalar mapping used to map component-specific parameters to the causal estimand; it does not require a fully linear potential outcome model. 

\subsection{AMUA properties and the two-component mixture representation}\label{ch2:sec:AMUA}

AMUA represents the assumption-hold and assumption-fail components within a single joint model,
\begin{align}
f_{\mathrm{AMUA}}(&\tilde{\mathbf z}_0,\tilde{\mathbf z}_1,\tilde{\mathbf z},
\tilde{\kappa},\widetilde{\boldsymbol{\theta}},\tilde{\omega},\tilde{\mathbf a},
\mathbf z,\kappa,\boldsymbol{\theta},\omega,\mathbf a,\delta
\mid \mathbf X,\pi).
\label{ch2:eq:famua_def}
\end{align}
Here, $\widetilde{\boldsymbol{\theta}}$ denotes the latent parameter block introduced in \eqref{ch2:eq:latent_blocks}, including $\widetilde{\boldsymbol\beta}^{(1)}$ and $\widetilde{\boldsymbol\beta}^{(0)}$. The AMUA construction satisfies four properties:
\begin{enumerate}
\item[\textbf{(P1)}] \textbf{Independence of data and component label.} The observed data $(\mathbf z,\mathbf a)$ should not learn $\delta$ from the likelihood. Consequently, $P(\delta\mid \mathbf z,\mathbf a)$ should not be updated by the observed data.
\item[\textbf{(P2)}] \textbf{Marginal recovery of the intended components.} When marginalizing across all quantities denoted with a ``$\sim$'' we obtain $f_{\mathrm{obs}}$, and similarly we obtain $f_{\mathrm{mix}}$ when marginalizing across quantities without a ``$\sim$''.
\item[\textbf{(P3)}] \textbf{Assumptions hold in Component 1.} When $\delta=1$, the latent quantities should be equal to the observed-data quantities with probability one.
\item[\textbf{(P4)}] \textbf{Flexibility in Component 2.} When $\delta=0$, the assumption-fail component must remain broad enough to accommodate violations of any of A1--A4.
\end{enumerate}

We construct the joint model in \eqref{ch2:eq:famua_def} to satisfy these properties. When $\delta=1$, AMUA is defined as
\begin{align}
	&f_{\mathrm{AMUA}}(\tilde{\mathbf z},\tilde{\kappa},\widetilde{\boldsymbol{\theta}},\tilde{\omega},\tilde{\mathbf a},
	\mathbf z,\kappa,\boldsymbol{\theta},\omega,\mathbf a,\delta=1\mid \mathbf X,\pi)
	\label{ch2:eq:delta1_joint}\\
	&\qquad=
	f_{\mathrm{Dirac}}(\tilde{\mathbf z},\tilde{\kappa},\widetilde{\boldsymbol{\theta}},\tilde{\omega},\tilde{\mathbf a}\mid \mathbf z,\kappa,\boldsymbol{\theta},\omega,\mathbf a,\delta=1)\;
	f_{\mathrm{obs}}(\mathbf z,\kappa,\boldsymbol{\theta},\omega,\mathbf a\mid \mathbf X)\;
	f(\delta=1\mid \pi).
	\nonumber
\end{align}
In \eqref{ch2:eq:delta1_joint}, $f_{\mathrm{Dirac}}$ denotes a product of point-mass distributions that forces the latent quantities to equal their observed-data counterparts:
\begin{align}
	&f_{\mathrm{Dirac}}(\tilde{\mathbf z},\tilde{\kappa},\widetilde{\boldsymbol{\theta}},\tilde{\omega},\tilde{\mathbf a}\mid \mathbf z,\kappa,\boldsymbol{\theta},\omega,\mathbf a,\delta=1)\nonumber\\
	&=
	f_{\mathrm{Dirac}}(\tilde{\mathbf z}-\mathbf z)
	f_{\mathrm{Dirac}}(\tilde{\kappa}-\kappa)
	f_{\mathrm{Dirac}}(\widetilde{\boldsymbol{\theta}}-\boldsymbol{\theta})
	f_{\mathrm{Dirac}}(\tilde{\omega}-\omega)
	f_{\mathrm{Dirac}}(\tilde{\mathbf a}-\mathbf a).
	\label{ch2:eq:dirac_factor}
\end{align}

\noindent
 In this component, the Dirac factor sets the latent quantities equal to their observed-data counterparts with probability one. In particular, \(f_{\mathrm{Dirac}}(\tilde{\mathbf z}-\mathbf z)\) places unit mass on \(\tilde{\mathbf z}=\mathbf z\), while the corresponding Dirac factors impose the analogous constraints for the remaining latent quantities. The observed-data factor \(f_{\mathrm{obs}}(\mathbf z,\kappa,\boldsymbol{\theta},\omega,\mathbf a\mid\mathbf X)\) contains no \(\delta\), so the observed-data likelihood is identical across components. Thus, when \(\delta=1\), the model reduces to the standard observed-data causal model, with the latent quantities constrained to satisfy the corresponding assumption-hold relationships.

When $\delta=0$, AMUA is defined as
\begin{align}
	&f_{\mathrm{AMUA}}(\tilde{\mathbf z},\tilde{\kappa},\widetilde{\boldsymbol{\theta}},\tilde{\omega},\tilde{\mathbf a},
	\mathbf z,\kappa,\boldsymbol{\theta},\omega,\mathbf a,\delta=0\mid \mathbf X,\pi)	\nonumber\\
	&\qquad=
	f_{\mathrm{mix}}(\tilde{\mathbf z},\tilde{\kappa},\widetilde{\boldsymbol{\theta}},\tilde{\omega},\tilde{\mathbf a},\delta=0\mid \mathbf X,\pi)
	\;f_{\mathrm{obs}}(\mathbf z,\kappa,\boldsymbol{\theta},\omega,\mathbf a\mid \mathbf X).
\label{ch2:eq:delta0_joint}
\end{align}
Under this specification, the marginal observed-data likelihood is
\begin{align}
	f_{\mathrm{AMUA}}(\mathbf z,\mathbf a\mid \mathbf X,\pi)
	=
	f_{\mathrm{obs}}(\mathbf a,\mathbf z\mid \mathbf X),
	\label{ch2:eq:marginal_data_same}
\end{align}
so the data likelihood does not identify $\pi$ again enforcing P1. Thus, sensitivity to the unchecked assumptions can be studied by varying \(\pi\) directly. Equations \eqref{ch2:eq:delta1_joint} and \eqref{ch2:eq:delta0_joint} thus imply the following component-specific marginals:
\begin{align}
\hspace{2pt}&f_{\mathrm{AMUA}}(\mathbf z,\kappa,\boldsymbol{\theta},\omega,\mathbf a,\delta=1\mid \mathbf X,\pi)
=
f_{\mathrm{obs}}(\mathbf z,\kappa,\boldsymbol{\theta},\omega,\mathbf a\mid \mathbf X)\,
f(\delta=1\mid \pi),
\label{ch2:eq:obs_marginal}
\\
\hspace{2pt}&f_{\mathrm{AMUA}}(\tilde{\mathbf z},
\tilde{\kappa},\widetilde{\boldsymbol{\theta}},\tilde{\omega},\tilde{\mathbf a},
\delta=0\mid \mathbf X,\pi)
 =
f_{\mathrm{mix}}(\tilde{\mathbf z},
\tilde{\kappa},\widetilde{\boldsymbol{\theta}},\tilde{\omega},\tilde{\mathbf a},
\delta=0\mid \mathbf X,\pi).
\label{ch2:eq:mix_marginal}
\end{align}
Here, $f_{\mathrm{obs}}(\cdot)$ refers to the Component 1 observational model in \eqref{ch2:eq:obs_joint_generic} and its marginal recovery in \eqref{ch2:eq:obs_marginal}. Thus, when \(\delta=1\), AMUA recovers the standard observational component together with its prior mass, whereas when \(\delta=0\), it recovers the flexible latent component used to represent possible assumption failures. The posterior distribution of $\widetilde{\boldsymbol{\theta}}$ thus takes the mixture form
\begin{align}
	p_{\mathrm{AMUA}}(\widetilde{\boldsymbol{\theta}}\mid \mathbf z,\mathbf a,\mathbf X,\pi)
	=
	\pi\,p_{\mathrm{obs}}(\widetilde{\boldsymbol{\theta}}\mid \mathbf z,\mathbf a,\mathbf X)
	+
	(1-\pi)\,p_{\mathrm{mix}}(\widetilde{\boldsymbol{\theta}}\mid \mathbf X,\delta=0).
	\label{ch2:eq:post_mix_theta}
\end{align}
Thus $\pi$ controls how much posterior mass is assigned to inference under the identifying assumptions. 

This specification of AMUA satisfies P1--P4. P1 is reflected in \eqref{ch2:eq:marginal_data_same}, because the observed-data likelihood is the same regardless of $\pi$ and therefore does not learn the component label. 
P2 is reflected in Equations \eqref{ch2:eq:obs_marginal} and \eqref{ch2:eq:mix_marginal}, which recover the two intended marginals. P3 is implemented by the Dirac factor in \eqref{ch2:eq:dirac_factor}, together with the Component 1 causal model under A1--A4. P4 is implemented by leaving $f_{\mathrm{mix}}$ unspecified except for its prior distributions on $\tilde{\bm{\beta}}^{(1)}$ and $\tilde{\bm{\beta}}^{(0)}$, which is needed to compute $\mathrm{ATE}_{\mathrm{CS}}(\tilde{\bm{\beta}}^{(1)},\tilde{\bm{\beta}}^{(0)},0)$.

\subsection{Hierarchical specification of Component 1 under AMUA}
\label{ch2:subsec:comp1spec}

\hspace{5mm}Although Component 1 was introduced in the general observational form in \eqref{ch2:eq:obs_blocks}--\eqref{ch2:eq:obs_joint_generic}, we now provide a more detailed hierarchical specification, which accommodates both the Gaussian response model used in the simulation study and the Poisson log-link model used in the application. For each unit $i$, let $\mu_i=\mathbb{E}(Z_i\mid A_i,\mathbf x_i,\boldsymbol\eta,\boldsymbol{\theta})$ denote the conditional mean of the observed response. We write
\begin{align*}
Z_i\mid \mu_i,\phi
\sim
{EF}(\mu_i,\phi),\ g(\mu_i)
=
\mathbf{x}_i(A_i)^\top \boldsymbol{\beta}
+
\mathbf g_i^\top \boldsymbol{\eta},
\end{align*}
where ${EF}(\mu_i,\phi)$ denotes an exponential-family distribution with mean $\mu_i$ and dispersion or scale parameter $\phi$, and $g(\cdot)$ is an appropriate link function. The first term represents the treatment-specific regression contribution, while the second represents the random-effect or spatial component. The Gaussian simulation model follows from an identity link and Gaussian likelihood, whereas the application uses a Poisson likelihood with a log link.

The treatment-specific stacked design is defined as
\begin{align*}
\mathbf{x}_i(A_i)
&=
\Big(\mathbf{x}_i^\top \mathbf 1\{A_i=1\},\ \mathbf{x}_i^\top \mathbf 1\{A_i=0\}\Big)^\top
\in \mathbb{R}^{2p},
\nonumber\\
\boldsymbol{\beta}
&=
\Big((\boldsymbol{\beta}^{(1)})^\top,(\boldsymbol{\beta}^{(0)})^\top\Big)^\top.
\end{align*}
{
At covariate value $\mathbf x_i$, the conditional mean difference is $\mathbf x_i^\top(\boldsymbol{\beta}^{(1)}-\boldsymbol{\beta}^{(0)})$, whose expectation over the covariate distribution yields the component-specific ATE in \eqref{ch2:eq:atecs_delta_def}--\eqref{ch2:eq:atecs0_beta_relation}. 
For the binary
specification, the treatment assignment mechanism is
represented by $f(\mathbf a\mid\mathbf X)$, so $\omega=\emptyset$, and
therefore the parameter blocks are defined as
}
\[
\boldsymbol{\theta}=\{\boldsymbol\beta,\sigma_\beta^2,\mathbf\Sigma_\eta,\sigma_\xi^2\ ,\sigma_\epsilon^2\},
\qquad
\boldsymbol\kappa=\{\boldsymbol\eta,\xi_1,\ldots,\xi_n\}.
\]
The prior distributions are
\begin{align*}
\boldsymbol{\beta} &\sim \mathcal{N}_{2p}(\mathbf 0_{2p},\sigma_\beta^2 I_{2p}),
&
\boldsymbol\eta &\sim \mathcal{N}_r(\mathbf 0_r,\mathbf\Sigma_\eta),
\nonumber\\
\sigma_\beta^2 &\sim \mathrm{IG}(a_\beta,b_\beta),
&
\sigma_\xi^2 &\sim \mathrm{IG}(a_\xi,b_\xi),
\nonumber\\
\sigma_\epsilon^2 &\sim \mathrm{IG}(a_\epsilon,b_\epsilon),
&
\mathbf\Sigma_\eta &\sim \mathrm{IW}_r(\Psi,m).
\end{align*}
These priors define the Gaussian simulation specification. For brevity, the full conditional distributions used in the Gibbs sampler are provided in the Supplementary Materials.

\subsection{New inferential questions}

\hspace{5mm}Using the main ATE in \eqref{ch2:eq:ate_main} and the component-specific contrast in \eqref{ch2:eq:atecs_delta_def}, the AMUA posterior can be summarized through draws of $\mathrm{ATE}_{\mathrm{CS}}$ induced by the posterior distribution of $\widetilde{\boldsymbol{\theta}}$. For any measurable set $B$, this posterior-predictive step can be written as
\begin{align}
P\{\mathrm{ATE}_{\mathrm{CS}}\in B\mid \mathbf z,\mathbf a,\mathbf X\}
=\int \mathbf 1\{\mathrm{ATE}_{\mathrm{CS}}(\widetilde{\boldsymbol{\theta}})\in B\}
\,p(\widetilde{\boldsymbol{\theta}}\mid \mathbf z,\mathbf a,\mathbf X)\,d\widetilde{\boldsymbol{\theta}}.
\label{ch2:eq:atecs_predictive_step}
\end{align}
Substituting the mixture posterior in \eqref{ch2:eq:post_mix_theta} into the posterior-predictive representation in \eqref{ch2:eq:atecs_predictive_step} yields Equation \eqref{ch2:eq:ATE_post_mix_general}.
\begin{align}
p_{\mathrm{AMUA}}(\mathrm{ATE}_{\mathrm{CS}}\mid \mathbf z,\mathbf a,\mathbf X,\pi)
={}&
\pi\,p_{\mathrm{obs}}\{\mathrm{ATE}_{\mathrm{CS}}(\widetilde{\boldsymbol\beta}^{(1)},\widetilde{\boldsymbol\beta}^{(0)},1)\mid \mathbf z,\mathbf a,\mathbf X\}\nonumber\\
&+
(1-\pi)\,p_{\mathrm{mix}}\{\mathrm{ATE}_{\mathrm{CS}}(\widetilde{\boldsymbol\beta}^{(1)},\widetilde{\boldsymbol\beta}^{(0)},0)\mid \mathbf X\}.
\label{ch2:eq:ATE_post_mix_general}
\end{align}
The component indicator $\delta$ is marginalized in this mixture, so we write the resulting quantity simply as $\mathrm{ATE}_{\mathrm{CS}}$. Thus, posterior uncertainty in the ATE reflects both uncertainty within each component and uncertainty about the validity of A1--A4 through the prior mixture weight $\pi$. This leads naturally to a diagnostic question: what is the smallest analyst-specified prior probability that assumptions A1--A4 all hold for which the HPD set for the ATE excludes zero? We denote this threshold diagnostic by $\pi_{\min}$.
 One practical procedure for computing $\pi_{\min}$ is:
\begin{enumerate}
\item Choose a grid of candidate values $\pi_1,\ldots,\pi_B$.
\item For each $\pi_b$, generate draws from $p_{\mathrm{AMUA}}(\mathrm{ATE}_{\mathrm{CS}}\mid \mathbf z,\mathbf a,\mathbf X,\pi_b)$.
\item For each $\pi_b$, compute the corresponding $95\%$ HPD set, denoted $\mathrm{HPD}_b$.
\item Calculate
$
\pi_{\min}
=\min\{\pi_b:\ 0\notin \mathrm{HPD}_b\}$.
\end{enumerate}

The value of $\pi_{\min}$ should be interpreted as a diagnostic. It allows the analyst to compare their prior belief that A1--A4 jointly hold with the minimum belief required for the posterior HPD set to exclude zero. AMUA therefore focuses on how the causal conclusion depends on uncertainty about the identifying assumptions rather than on improving point estimation of the ATE.

Another inferential question that we answer is the following: What is the implicit prior probability of A1--A4 for an existing model that aims to address unchecked assumptions? To do this one can choose the value of $\pi$ so that the posterior density of AMUA is ``close'' to the posterior density of an existing model. For example, consider a spatial statistical model $f_{\mathrm{spatial}}(\mathrm{ATE}\vert \textbf{z},\textbf{a},\textbf{X})$, and a new diagnostic,
\[
\pi^\star
=
\arg\min_{\pi}
D\!\left\{
f_{\mathrm{AMUA}}(\,\cdot \vert \cdot,\pi),
f_{\mathrm{spatial}}
\right\},
\]
where $D$ is a specified distance or divergence. We use integrated squared error (ISE) for this purpose, with additional technical details provided in the Supplementary materials. The resulting $\pi^\star$ produces a posterior closest to that of the spatial model, providing a diagnostic for the prior belief in the joint validity of A1--A4 implicitly represented by the existing model.


\section{Simulation Study}
\label{ch2:sec:simulation}

\subsection{Simulation setup}

This simulation study has two goals. First, we evaluate AMUA when latent ignorability (A3) is violated by unmeasured confounding. Second, we illustrate the computation and interpretation of the diagnostic $\pi_{\min}$. We consider the binary-treatment case, $A_i\in\{0,1\}$.

To induce violation of A3, we introduce an unmeasured confounder $U_i$. For $i=1,\dots,n$, the data is generated as follows: 
\begin{align*} 
Z_i &= Y_i + \epsilon_i, \qquad \epsilon_i \sim \mathcal{N}(0,\sigma_\epsilon^2),\\ 
Y_i &= \mathbf{x}_i(A_i)^\top \boldsymbol{\beta} + \mathbf{g}_i^\top \boldsymbol{\eta} + U_i + \xi_i,\qquad \xi_i \sim \mathcal{N}(0,\sigma_\xi^2),\\ 
A_i &\mid \boldsymbol{\alpha}, \mathbf{x}_i, U_i \sim \mathrm{Bernoulli}\!\left(\mathrm{expit}(\mathbf{x}_i^\top\boldsymbol{\alpha} + \lambda U_i)\right),\\ 
U_i &= \mathbf{x}_i^\top\boldsymbol{\gamma} + \nu_i,\qquad \nu_i \sim \mathcal{N}(0,\sigma_U^2),
\end{align*}
Here, $\mathrm{expit}(t) = (1+\exp(-t))^{-1}$, the covariate vector is $\mathbf{x}_i=(1,x_{1i},x_{2i})^\top$, where $(x_{1i},x_{2i})$ is generated from a uniform grid on $[0,1]\times[0,1]$. The vectors $\boldsymbol{\alpha}$ and $\boldsymbol{\gamma}$ determine the effects of $\mathbf{x}_i$ on treatment assignment and the unmeasured confounder, respectively, while $\lambda$ controls the strength of the association between $U_i$ and treatment assignment. Because $U_i$ affects both the outcome and treatment assignment and is unobserved by the analyst, A3 is violated. We focus on unmeasured confounding because it is a central unchecked assumption in observational studies and is directly tied to A3. Other failures, such as spillover or interference, may also be possible; in spatial settings, unmeasured spatial confounding and spillover can be closely related sources of bias \citep{ReichYangGuan2021}.

We consider sample sizes
$n\in\{400,900,1600,2500\}$ and generate 200 replicated datasets for each scenario. The strength of confounding is varied through the magnitude of $\boldsymbol{\gamma}$, with values ranging from 2 to 7, and we examine how the conventional analysis corresponding to $\pi=1$ is affected as confounding increases. We set
$
\boldsymbol{\beta}_{\mathrm{true}}^{(1)}
=(2,4,2)^\top$
and
$\boldsymbol{\beta}_{\mathrm{true}}^{(0)}
=(1,2,1)^\top,
$
so that
\[
\mathrm{ATE}_{\mathrm{true}}
=
(1,0.5,0.5)
\left\{
(2,4,2)^\top-(1,2,1)^\top
\right\}
=2.5.
\]

For each replicated dataset, we evaluate the AMUA posterior over the prespecified grid
$\pi_b\in\{0.50,0.51,\ldots,0.99,1.00\}.
$
For each $\pi_b$, we compute the 95\% HPD set for the ATE and define $\pi_{\min}$ as the smallest grid value for which the HPD set excludes zero.

To sample from the AMUA posterior, we first draw
$
\delta^{(m)}\sim\mathrm{Bernoulli}(\pi)$.
When $\delta^{(m)}=1$, we draw the ATE from the posterior of the observed-data model. When $\delta^{(m)}=0$, we generate $\mathrm{ATE}_{\mathrm{CS}}$ from the assumption-fail component. In the latter case, we sample
$$
\sigma_\beta^2\sim\mathrm{IG}(a_\beta,b_\beta),
\qquad
\boldsymbol{\beta}\mid\sigma_\beta^2
\sim
\mathcal{N}(\mathbf{0},\sigma_\beta^2\mathbf{I}),
$$
and evaluate $\mathrm{ATE}_{\mathrm{CS}}$ as the corresponding linear functional of $\boldsymbol{\beta}$. Integrating out $\sigma_\beta^2$ gives
$
\boldsymbol{\beta}
\sim
t_{2a_\beta}
\left(
\mathbf{0},
\frac{b_\beta}{a_\beta}\mathbf{I}
\right).
$

\subsection{Results}

\begin{figure}[!t]
	\centering
	\includegraphics[width=0.80\textwidth]{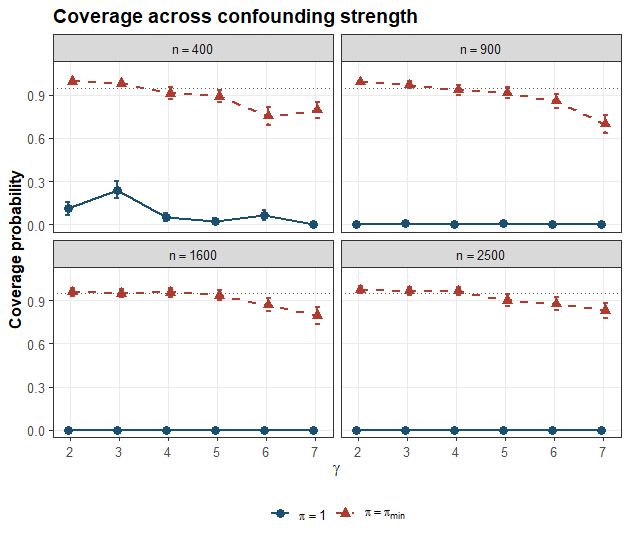}
	\caption{Coverage probability over 200 replicated datasets. Solid lines correspond to $\pi=1$, dashed lines correspond to $\pi_{\min}$, and the dotted line marks nominal 95\% coverage. Error bars show approximate 95\% Monte Carlo intervals.}
	\label{ch2:fig:sim_coverage}
\end{figure}
\begin{figure}[!t]
\centering
\includegraphics[width=0.70\textwidth]{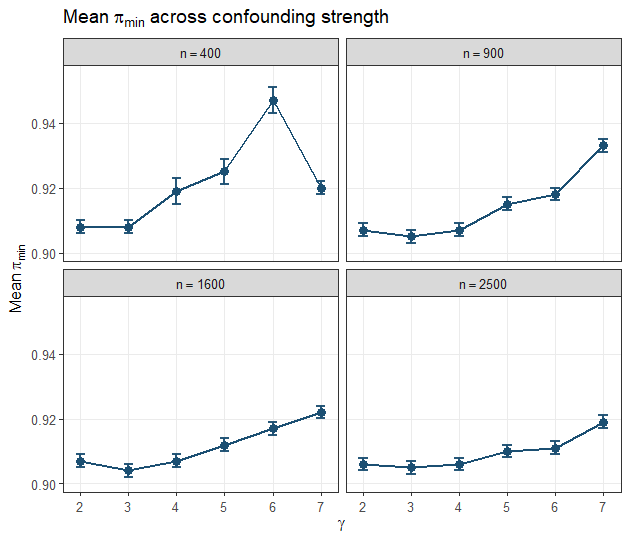}
\caption{Mean $\pi_{\min}$ over 200 replicated datasets. Panels correspond to sample sizes, and the horizontal axis gives the unmeasured-confounding strength $\gamma$. Error bars show approximate 95\% Monte Carlo intervals.}
\label{ch2:fig:sim_pimin}
\end{figure}
\begin{figure}[!htbp]
\centering
\includegraphics[width=0.79\textwidth]{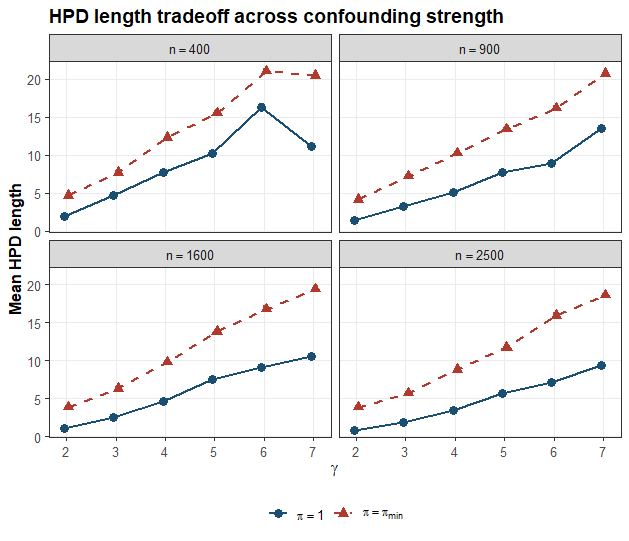}
\caption{Mean 95\% HPD length over 200 replicated datasets. Solid lines correspond to $\pi=1$, and dashed lines correspond to $\pi_{\min}$. Error bars show approximate 95\% Monte Carlo intervals.}
\label{ch2:fig:sim_hpdlen}
\end{figure}
\begin{figure}[!htbp]
\centering
\begin{tabular}{cc}
\includegraphics[width=0.48\textwidth]{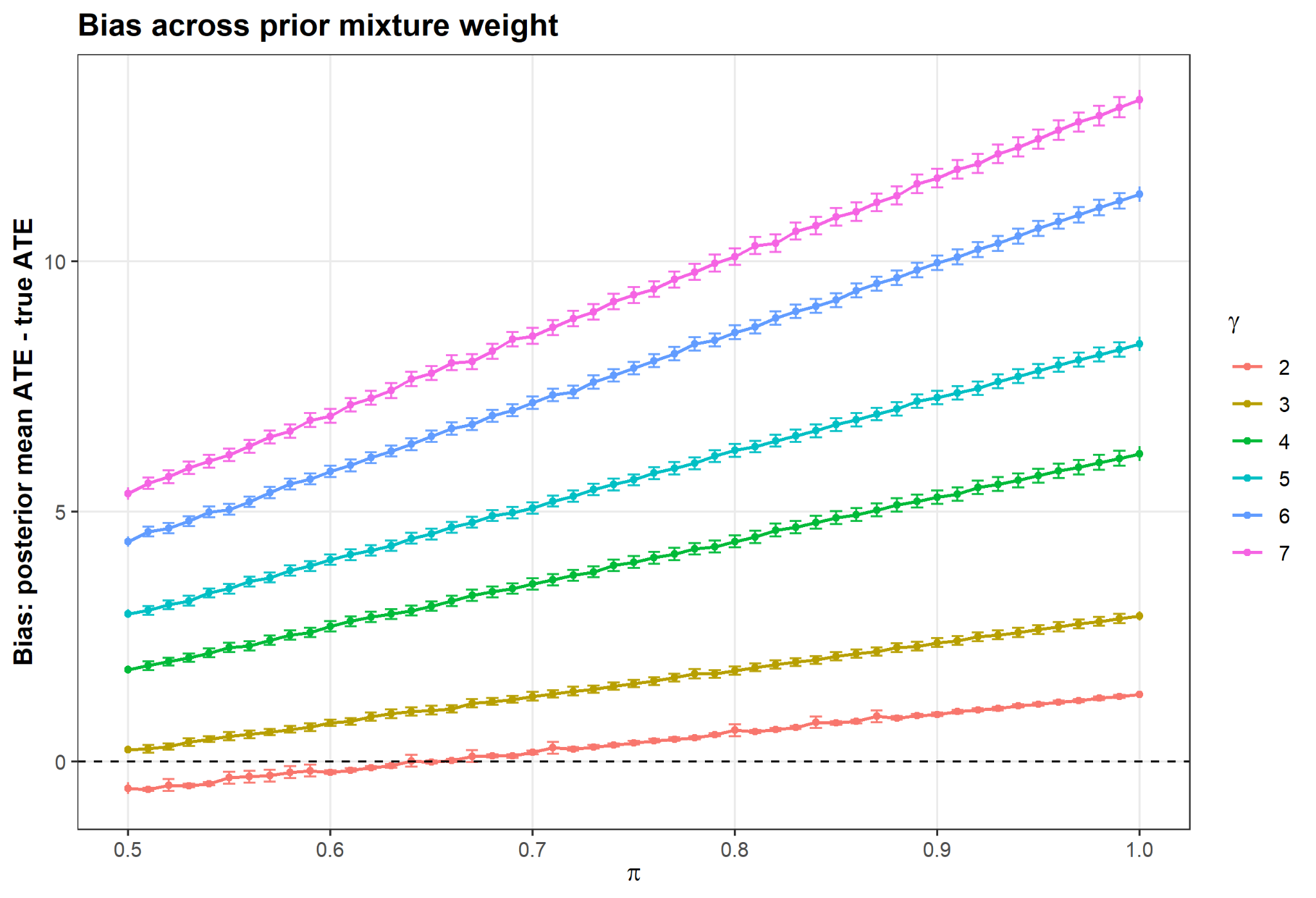}
&
\includegraphics[width=0.48\textwidth]{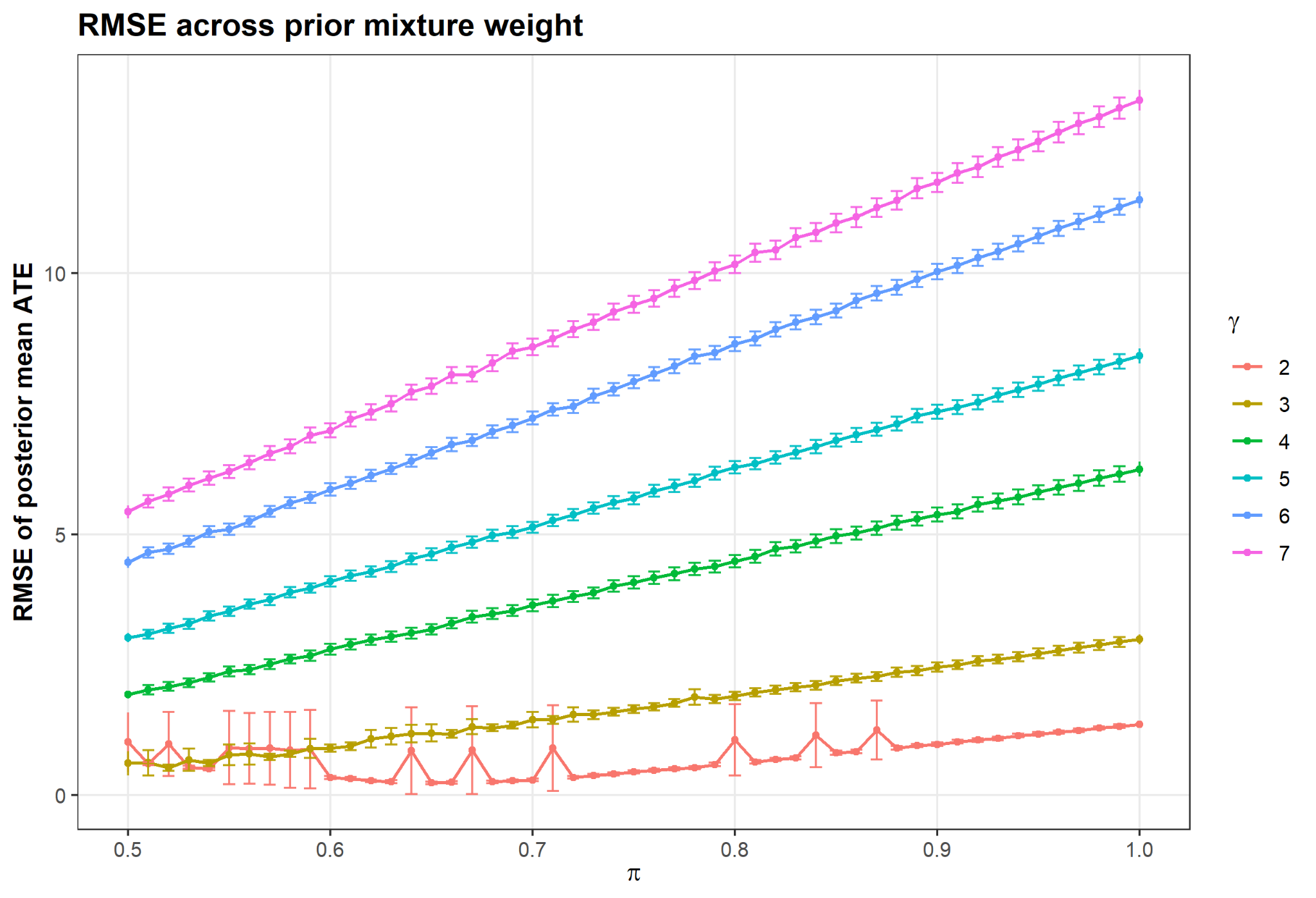}
\end{tabular}
\caption{Bias and RMSE of the posterior mean ATE across $\pi$ for $n=400$ and $\lambda=1$. The left panel shows bias, and the right panel shows RMSE. Curves correspond to different values of $\gamma$; error bars show approximate 95\% Monte Carlo intervals.}
\label{ch2:fig:sim_bias_rmse}
\end{figure}

The following metrics are evaluated across the 200 replicated datasets. We report the $95\%$ HPD coverage probability at $\pi=1$ and $\pi_{\min}$, defined as the proportion of datasets in which the corresponding $95\%$ HPD interval covers the true ATE; the mean $\pi_{\min}$ across datasets; the mean length of the $95\%$ HPD intervals at $\pi=1$ and $\pi_{\min}$; and the bias and RMSE of the posterior mean of the ATE across $\pi$ and $\gamma$. Although AMUA is primarily designed to provide the diagnostics $\pi_{\min}$ and $\pi^*$, we also examine its effect on bias and RMSE. Monte Carlo uncertainty is reported for all metrics, with the corresponding results shown in Figures~\ref{ch2:fig:sim_coverage}--\ref{ch2:fig:sim_bias_rmse}.

We now summarize the main patterns in Figures~\ref{ch2:fig:sim_coverage}--\ref{ch2:fig:sim_bias_rmse}. Figure~\ref{ch2:fig:sim_coverage} compares $95\%$ HPD coverage at $\pi=1$ and $\pi_{\min}$. Under stronger confounding, coverage at $\pi=1$ can become very low, in some settings approaching zero as $n$ and $\gamma$ increase. In contrast, coverage at $\pi_{\min}$ is substantially higher across the scenarios considered. Figure~\ref{ch2:fig:sim_pimin} shows the mean $\pi_{\min}$ across the 200 replicates. The mean is typically close to $0.90$ and tends to increase with $\gamma$, although the pattern is less consistent for $n=400$. For some settings, the threshold decreases slightly with increasing $n$, suggesting greater stability of the posterior ATE summary; these patterns should be viewed as simulation-specific rather than theoretical guarantees.

Figure~\ref{ch2:fig:sim_hpdlen} shows the corresponding interval lengths. The $95\%$ HPD intervals at $\pi_{\min}$ are consistently longer than those at $\pi=1$, reflecting the additional posterior uncertainty from the assumption-fail component. Although the $\pi=1$ intervals become shorter as $n$ increases, the corresponding coverage can deteriorate under strong confounding. Thus, the longer intervals at $\pi_{\min}$ reflect the additional uncertainty propagated by AMUA rather than a loss of precision under the assumed model.

Finally, Figure~\ref{ch2:fig:sim_bias_rmse} shows bias and RMSE of the posterior mean ATE across $\pi$ and $\gamma$ for the $n=400$, $\lambda=1$ setting. Bias and RMSE generally increase with confounding strength and with $\pi$ approaching one, although the patterns are not necessarily monotone. Lower values of $\pi$ can reduce bias and RMSE in some settings with strong confounding, as greater posterior weight is assigned to the assumption-fail component. These improvements are setting-specific, but become more pronounced in our simulations as the violation of A3 increases.


\section{Application}\label{ch2:sec:application}

\subsection{Setup}\label{ch2:subsec:app-data}

We analyze the U.S.\ county-level data from \citet{ReichYangGuan2021}. The analytic sample contains approximately $n=3{,}109$ counties, with county-level exposure, outcome, and covariate information. The response $Y_i$ is the cumulative number of COVID-19 deaths in county $i$, with $\log N_i$ included as an offset, where $N_i$ is the county population. The exposure $A_i$ is average PM$_{2.5}$ concentration, measured in $\mu\mathrm{g}/\mathrm{m}^3$, and $\mathbf{x}_i\in\mathbb{R}^{15}$ contains socio-demographic and health-related covariates selected to capture potential confounding.

Our primary interest is the effect of long-term PM$_{2.5}$ exposure on COVID-19 mortality. Under the Poisson regression model with a log link, the PM$_{2.5}$ effect is parameterized by the coefficient $\beta_A$, which corresponds to the ATE for a one-unit increase in the continuous exposure under the specified model (see the Supplementary Materials for details). Specifically, a $1\,\mu\mathrm{g}/\mathrm{m}^3$ increase in PM$_{2.5}$ corresponds to a relative risk of
\[
RR=\exp(\beta_A).
\]
Thus, $\beta_A>0$ indicates increased mortality with higher PM$_{2.5}$ exposure, whereas $\beta_A<0$ indicates decreased mortality.

We apply AMUA to propagate uncertainty about the causal identifying assumptions A1--A4 through the mixture weight $\pi\in[0,1]$. When $\pi=1$, all assumptions are treated as holding and inference reduces to the baseline observational model; when $\pi<1$, the assumption-fail component receives positive weight. We examine how the $95\%$ HPD interval for the treatment effect changes with $\pi$ and report $\pi_{\min}$, the smallest value of $\pi$ for which the interval excludes zero.

\subsection{The Observational Model and the AMUA Specifications}\label{ch2:subsec:app-models}

We use the two-stage spatial model of \citet{ReichYangGuan2021}, with separate exposure and outcome models. In Stage 1, exposure $A_i$ is modeled as a function of observed covariates and a spatial random effect, and a residual-based summary is constructed to characterize the exposure assignment mechanism. In Stage 2, mortality is modeled as a function of PM$_{2.5}$, covariates, a spatial random effect, and a spline function of this summary to account for residual confounding under continuous exposure. Covariate preprocessing was performed before model fitting. After removing variables with excessive missingness, continuous covariates were standardized to mean zero and variance one, and an intercept was included in the design matrix. Spatial adjacency was represented by a first-order symmetric binary matrix $\boldsymbol W=\{w_{ij}\}$, where $w_{ij}=1$ if counties $i$ and $j$ are neighbors and $w_{ij}=0$ otherwise. Let $\boldsymbol D=\mathrm{diag}(d_1,\ldots,d_n)$, where $d_i=\sum_{j=1}^n w_{ij}$.

The Stage 1 exposure model is
\[
A_i \;=\; \mathbf{x}_i^\top \boldsymbol{\alpha} \;+\; V_i \;+\; \epsilon_i,
\qquad 
\epsilon_i \sim \mathcal{N}(0,\sigma_{\epsilon}^2),
\]
where $\mathbf{V}=(V_1,\ldots,V_n)^\top$ follows a proper CAR prior:
\[
p(\mathbf{V} \mid \rho_V,\tau_V) \;\propto\; 
\exp\!\Big(-\frac{\tau_V}{2}\, \mathbf{V}^\top \big[\boldsymbol{D}-\rho_V \boldsymbol{W}\big] \mathbf{V}\Big),
\qquad 
\rho_V\sim\mathrm{Unif}(0,1),\ \tau_V>0.
\]
The spatial effect $V_i$ captures spatial structure in exposure not explained by observed covariates. Following \citet{ReichYangGuan2021}, we use the squared residual
\[
\hat e_i \;=\; \big(A_i - (\mathbf{x}_i^\top \boldsymbol{\alpha} + V_i)\big)^2,
\]
as a GPS summary and construct a $K=5$ dimensional B-spline basis $\boldsymbol b(\hat e_i)$ for flexible adjustment in Stage 2.

The Stage 2 outcome model is
\[
Y_i \mid \mu_i \sim \mathrm{Poisson}(\mu_i), \qquad
\log \mu_i
=
\log N_i
+
A_i \beta_A
+
\mathbf{x}_i^\top \boldsymbol{\gamma}
+
U_i
+
\boldsymbol{b}(\hat e_i)^\top \boldsymbol{\theta},
\]
where $\mu_i$ is the expected where $N_i$ is the county population and $\mathbf U=(U_1,\ldots,U_n)^\top$ follows the proper CAR prior
\[
p(\mathbf{U} \mid \rho_U,\tau_U) \;\propto\; 
\exp\!\Big(-\frac{\tau_U}{2}\, \mathbf{U}^\top \big[\boldsymbol{D}-\rho_U \boldsymbol{W}\big] \mathbf{U}\Big),
\qquad 
\rho_U\sim\mathrm{Unif}(0,1),\ \tau_U>0.
\]
Equivalently, with $\lambda_i=\mu_i/N_i$ denoting the county-specific mortality rate,
\[
\log\lambda_i
=
A_i\beta_A
+
\mathbf{x}_i^\top\boldsymbol{\gamma}
+
U_i
+
\boldsymbol b(\hat e_i)^\top\boldsymbol\theta.
\]
Here, $\beta_A$ is the PM$_{2.5}$ effect of interest, $\mathbf{x}_i^\top\boldsymbol{\gamma}$ adjusts for observed covariates, $U_i$ captures residual spatial structure, and the spline term provides flexible adjustment for the exposure assignment mechanism. The GPS spline term is included only in the assumption-hold component, where $\boldsymbol\theta$ is estimated from the data. In the assumption-fail component, its coefficients are fixed at zero, so the component-specific exposure effect is represented by $\widetilde{\beta}_A$ without GPS spline adjustment. This allows the AMUA mixture to distinguish inference under the assumed causal model from inference that does not rely on the GPS adjustment. After fitting the assumption-hold model, we apply AMUA to posterior draws of $\beta_A$. For this continuous-exposure application, the $\pi_{\min}$ diagnostic is based on the $95\%$ HPD set for $\beta_A$, corresponding to the log-relative-risk contrast for a one-unit increase in PM$_{2.5}$.

We use weakly informative regression priors and proper CAR priors for the spatial random effects, following the prior structure of \citet{ReichYangGuan2021}. Posterior computation is performed using OpenBUGS through \texttt{R2OpenBUGS} and Stan. The main paper reports posterior summaries for $\beta_A$ and its AMUA $\pi$-sensitivity summaries; detailed prior specifications, MCMC settings, and convergence diagnostics are provided in the Supplementary materials.

\subsection{Results}\label{ch2:subsec:app-results} 

\begingroup

We compared four models considered in \citet{ReichYangGuan2021}. The non-spatial model without the
GPS spline is denoted by NS, and its GPS-adjusted counterpart is denoted by
NS+P. The corresponding models with a dependent spatial random effect are
denoted by S and S+P, respectively. Thus, the suffix \(+\mathrm P\) indicates
inclusion of the GPS B-spline adjustment, whereas the contrast between NS and
S indicates whether dependent spatial structure is included in the outcome
random effect. We see more conservative results using a spatial model in Table~\ref{ch2:tab:app_betaA_model_comparison}. Thus, we consider $\pi^{*}$ introduced in Section 2.6 to help interpret these more conservative values. In fact, the non-spatial models suggest one needs to have a high prior belief on A1--A4 to determine a causal conclusion, with $\pi_{\mathrm{min}} = 0.95$ for NS and equal to 0.93 for NS+P.

\begin{table}[!htbp]
\centering
\small
\caption{Posterior summaries for the PM$_{2.5}$ coefficient \(\beta_A\) under
the four assumption-hold outcome-model variants.}
\label{ch2:tab:app_betaA_model_comparison}
\begin{tabular}{lrrlc}
\toprule
Model & Mean & SD & 95\% HPD & Zero included?\\
\midrule
NS   & 0.0762 & 0.0323 & \([0.0130,\,0.1382]\)  & No\\
NS+P & 0.0992 & 0.0330 & \([0.0386,\,0.1680]\)  & No\\
S    & 0.0332 & 0.0521 & \([-0.0688,\,0.1390]\) & Yes\\
S+P  & 0.0669 & 0.0553 & \([-0.0454,\,0.1715]\) & Yes\\
\bottomrule
\end{tabular}
\end{table}

Table~\ref{ch2:tab:app_betaA_model_comparison} shows that the NS and NS+P HPD
intervals exclude zero, whereas the S and S+P intervals include zero.
Accordingly, the posterior conclusion changes materially when dependent
spatial structure is introduced. The GPS spline increases the posterior mean
within both the non-spatial and spatial pairs, but its effect is distinct from
the increase in posterior uncertainty associated with spatial adjustment.

To specify $\pi^{*}$ we define an empirically defined assumption-fail component. Specifically, let 
\[
\beta_A\mid\delta = 0,m,c
\sim N\!\left(0,c^{2}s_{m}^2\right),
\qquad
m\in \{\mathrm{NS},\mathrm{NS}+\mathrm{P}\},
\]
where \(c\) is a scale multiplier controlling how diffuse the assumption-fail
component is relative to the posterior standard deviation, denoted $s_{m}^{2}$ from the assumption-hold component. We considered
\(c\in\{1.00,1.25,\ldots,3.00\}\). Here, \(c=1\) forces the assumption-fail component to match the assumption-hold posterior standard deviation, whereas larger values allow progressively greater uncertainty under the
assumption-fail component. Again, $\pi^{*}$ is the value of $\pi$ that minimizes the distance between the density under the AMUA model with the assumption-fail component defined above, and that of the model with a spatial random-effect. To do this, we select $\pi$ and $c$ by minimizing the ISE. The selected values of were $c^\star=1.25$ and $\pi^{*} = 0.41$ for NS, and $c^\star=1.50$ and $\pi^{*} = 0.56$ for NS+P. 

\begin{figure}[!htbp]
\centering
\IfFileExists{Fig_selected_2panel_NS_1.25_NSP_1.50.png}{%
  \includegraphics[width=0.96\textwidth]
  {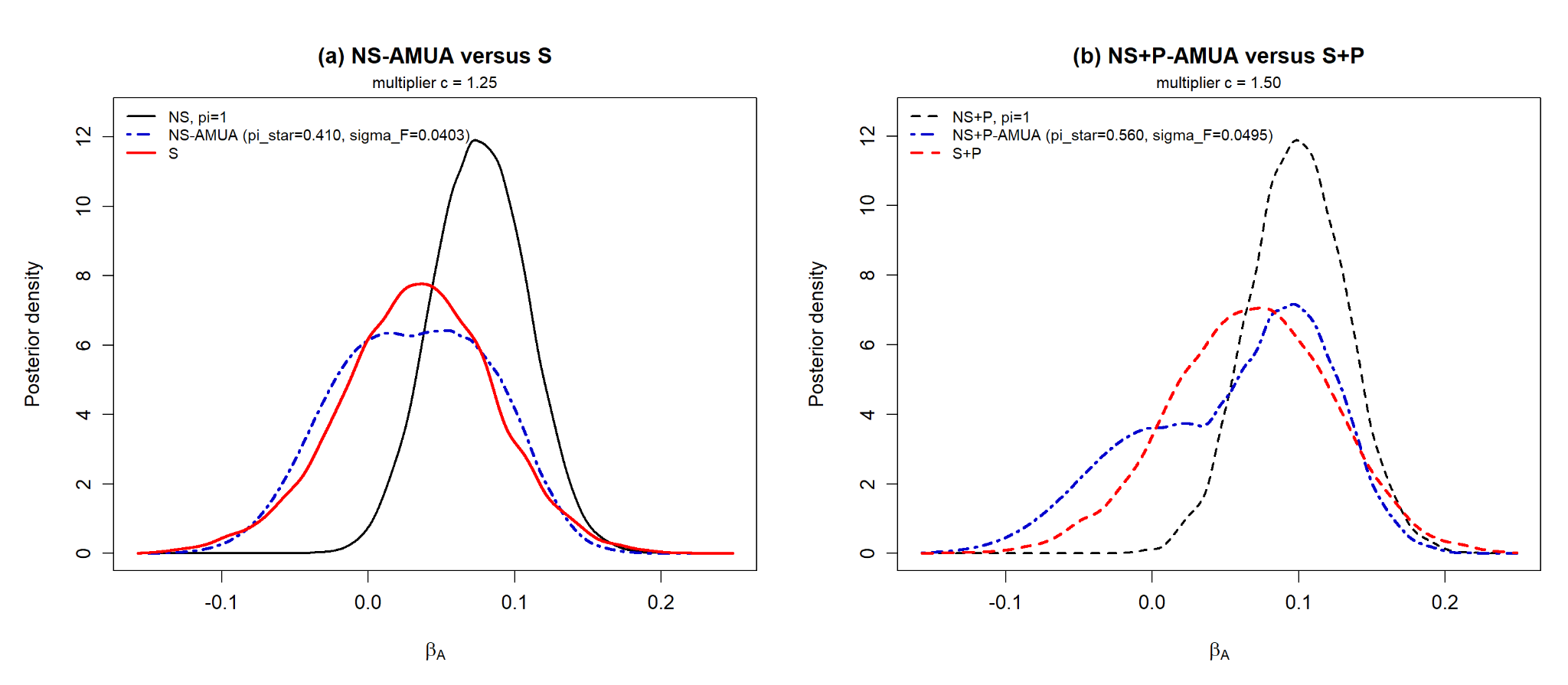}%
}{%
  \IfFileExists{Fig_selected_2panel_NS_1.25_NSP_1.50.png}{%
    \includegraphics[width=0.96\textwidth]
    {Fig_selected_2panel_NS_1.25_NSP_1.50.png}%
  }{%
    \fbox{\parbox{0.90\textwidth}{\centering
    Insert \texttt{\detokenize{Fig_selected_2panel_NS_1.25_NSP_1.50.png}}
    here.}}%
  }%
}
\caption{Posterior density under the selected baseline-specific
assumption-fail components. The left panel compares the NS posterior,
the NS-AMUA posterior at \(c^\star=1.25\) and \(\pi^\star=0.41\), and the S
posterior. The right panel compares the NS+P posterior, the NS+P-AMUA
posterior at \(c^\star=1.50\) and \(\pi^\star=0.56\), and the S+P posterior.
The selected AMUA distributions are the blue curves, the non-spatial
baselines are the black curves, and the corresponding spatial targets are the
red curves.}
\label{ch2:fig:app_selected_density_matching}
\end{figure}

At the selected $\pi^{*}$, the NS-AMUA HPD interval was
\([-0.0666,0.1289]\), and the NS+P-AMUA interval was
\([-0.0722,0.1600]\). Both intervals include zero, consistent with the
corresponding S and S+P posteriors. The calibrated values
$\pi^\star=0.41$ for NS and $\pi^\star=0.56$ for NS+P indicate that
the corresponding spatial-model posteriors are closest, under the ISE
criterion, to AMUA posteriors with these prior weights assigned to the
assumption-hold component. The larger $\pi^\star$ for NS+P suggests that
incorporating the propensity-score adjustment produces a spatial-model
posterior more similar to an AMUA analysis assigning greater prior weight
to the joint validity of A1--A4.
\endgroup


\section{Discussion}\label{ch2:sec:discussion}

This paper proposes the Adjusted Model for Unchecked Assumptions (AMUA), a Bayesian framework that propagates uncertainty about causal identifying assumptions that cannot be verified from observational data alone. AMUA represents inference through a two-component mixture comprising an assumption-hold component and an assumption-fail component. The mixing weight $\pi$ represents the analyst's prior probability that the identifying assumptions jointly hold. Lower values of $\pi$ assign greater weight to the assumption-fail component and thereby propagate additional uncertainty about the causal effect.

In simulations with unmeasured confounding, the $\pi=1$ analysis produced increasingly short HPD intervals and substantially lower coverage as confounding strengthened. In contrast, inference at $\pi_{\min}$ yielded higher coverage at the cost of longer intervals, illustrating the tradeoff between precision and uncertainty about assumption validity. In our simulation settings, lower values of $\pi$ also reduced bias and RMSE when A3 was violated, although these improvements were setting-specific. In the real-data application, the NS and NS+P assumption-hold components excluded zero, whereas the S and S+P posteriors included zero. The corresponding $\pi_{\min}$ values were $0.95$ for NS and $0.93$ for NS+P, indicating that relatively high prior probability of joint validity of A1--A4 was required for the corresponding credible sets to exclude zero. The $\pi^\star$ diagnostics were substantially smaller: $\pi^\star=0.41$ for NS-AMUA versus S and $\pi^\star=0.56$ for NS+P-AMUA versus S+P. Thus, the spatial models produced posterior distributions similar to AMUA analyses that assigned substantially less prior probability to joint validity of A1--A4. 

For reporting, we recommend that AMUA applications provide: (i) baseline results at $\pi=1$, including a point estimate and $95\%$ credible or HPD set; (ii) the assumption-fail specification and rationale for its prior scale; (iii) posterior results across a grid of $\pi$ values, including posterior means, interval limits, and interval lengths; and (iv) $\pi_{\min}$, defined as the smallest $\pi$ for which the chosen credible set excludes zero. When the $\pi=1$ credible set includes zero, $\pi^\star$ can additionally be used to assess how the posterior from an existing observational model compares with AMUA inference under different levels of prior belief in assumption validity.

\section*{Acknowledgements}
We thank Dr. Brian J. Reich (North Carolina State University) for kindly providing the county-level COVID-19 and PM$_{2.5}$ data used in our application. Jonathan Bradley's research is partially funded by NSF-DMS.

\section*{Data Availability}
The data analyzed in this study are publicly available in the repository accompanying \cite{WuNetherySabathBraunDominici2020}: \url{https://github.com/wxwx1993/PM_COVID}.

\bibliographystyle{plainnat}
\bibliography{refs}

\clearpage
\thispagestyle{plain}

\begin{center}
    {\Large\bfseries Supplementary Material\par}
    \vspace{1.5em}
    {\normalsize
    Supplementary Material for\\[0.5em]
    ``A Subjective Bayesian Approach to Address Unchecked Assumptions
    in a Causal Analysis''\\[0.5em]
    by Jaiyool Kim, Indrabati Bhattacharya, and Jonathan R. Bradley
    \par}
\end{center}

\vspace{1em}

\renewcommand{\thesection}{Appendix \Alph{section}}
\setcounter{section}{0}
\setcounter{table}{0}
\setcounter{figure}{0}
\renewcommand{\thefigure}{S\arabic{figure}}
\renewcommand{\thetable}{S\arabic{table}}

\renewcommand{\theHfigure}{supp.figure.\arabic{figure}}
\renewcommand{\theHtable}{supp.table.\arabic{table}}
\renewcommand{\theHsection}{supp.\Alph{section}}
\renewcommand{\theHtable}{supp.\arabic{table}}
\renewcommand{\theHfigure}{supp.\arabic{figure}}
\makeatletter
\makeatother

\section{The Gibbs Sampler for the Gaussian Observational Model} In this section, we outline the full conditional distributions for the Gaussian observational model. To derive the full conditional distribution for $\boldsymbol{\beta}$, we focus on the terms in the joint posterior density involving $\boldsymbol{\beta}$ as follows:

$$
\begin{aligned}
f\left(\boldsymbol{\beta}|\cdots\right) & \propto \exp \left\{-\frac{1}{2\cdot \sigma_{\epsilon}^2}\left(\boldsymbol{Z}-(\mathbf{X} \boldsymbol{\beta}+\mathbf{G} \boldsymbol{\eta}+\boldsymbol{\xi})\right)^{\top}(\boldsymbol{Z}-(\mathbf{X} \boldsymbol{\beta}+\mathbf{G} \boldsymbol{\eta}+\boldsymbol{\xi}))\right\} \times \exp \left\{-\frac{1}{2\cdot \sigma_{\beta}^2} \boldsymbol{\beta}^{\top} \boldsymbol{\beta}\right\} \\
& = \exp\Bigg\{-\frac{1}{2}\Bigg[\boldsymbol{\beta}^{\top}\left(\frac{1}{\sigma_{\epsilon}^2}\mathbf{X}^{\top} \mathbf{X}+\frac{1}{\sigma_{\beta}^2}\mathbf{I}_{2 p}\right) \boldsymbol{\beta}-2 \boldsymbol{\beta}^{\top} \left(\frac{1}{\sigma_{\epsilon}^2}\mathbf{X}^{\top}(\boldsymbol{Z}-\mathbf{G} \boldsymbol{\eta}-\boldsymbol{\xi})\right) \\
& + \frac{1}{\sigma_{\epsilon}^2} \boldsymbol{Z}^{\top} \boldsymbol{Z}-\frac{2}{\sigma_{\epsilon}^2}\left(\boldsymbol{\eta}^{\top} \mathbf{G}^{\top} +\boldsymbol{\xi}^{\top}\right) \boldsymbol{Z}+ \frac{1}{\sigma_{\epsilon}^2}\left(\boldsymbol{\eta}^{\top} \mathbf{G}^{\top}+\boldsymbol{\xi}^{\top}\right) \left(\mathbf{G} \boldsymbol{\eta} +\boldsymbol{\xi}\right)\Bigg]\Bigg\} \\
& \propto \exp\Bigg\{-\frac{1}{2}\Bigg[\boldsymbol{\beta}^{\top}\left(\frac{1}{\sigma_{\epsilon}^2}\mathbf{X}^{\top} \mathbf{X}+\frac{1}{\sigma_{\beta}^2}\mathbf{I}_{2 p}\right) \boldsymbol{\beta}-2 \boldsymbol{\beta}^{\top} \left(\frac{1}{\sigma_{\epsilon}^2}\mathbf{X}^{\top}(\boldsymbol{Z}-\mathbf{G} \boldsymbol{\eta}-\boldsymbol{\xi})\right)\Bigg]\Bigg\}. 
\end{aligned}
$$

\noindent It has the multivariate normal kernel as above. Therefore, the full conditional distribution of $\boldsymbol{\beta}$ is the multivariate normal distribution as follows:
\begin{itemize}
\item $\left[\boldsymbol{\beta} \mid \cdots \right] \sim \mathrm{N}_{2p}\left(\boldsymbol{\mu}_{\boldsymbol{\beta}}, \boldsymbol{\Sigma}_{\boldsymbol{\beta}}\right)$, \\ \hspace{0.5mm} where $\boldsymbol{\Sigma}_{\boldsymbol{\beta}}=\left(\frac{1}{\sigma_{\epsilon}^2}  \mathbf{X}^{\top}\mathbf{X}+\frac{1}{{\sigma_{\beta}^2}} \mathbf{I}_{2p}\right)^{-1},$ \\
$\boldsymbol{\mu}_{\boldsymbol{\beta}}=\boldsymbol{\Sigma}_{\boldsymbol{\beta}}\left(\frac{1}{\sigma_{\epsilon}^2} \mathbf{X}^{\top}\left(\boldsymbol{Z}-\mathbf{G}\boldsymbol{\eta}-\boldsymbol{\xi} \right) \right),$
\end{itemize}
where $\boldsymbol{\mu}_{\boldsymbol{\beta}}$ has a dimension $2p$ $\times$ $1$, and $\boldsymbol{\Sigma}_{\boldsymbol{\beta}}$ is in the positive definite matrix space $2p$ $\times$ $2p$.

The matrix \(\mathbf{X}^{\top}\mathbf{X}\) is symmetric and positive semi-definite, allowing for eigen decomposition as follows:
\[
\mathbf{X}^{\top}\mathbf{X} = \mathbf{P} \mathbf{\Lambda} \mathbf{P}^{\top},
\]
where:
\begin{itemize}
\item \(\mathbf{P}\) is an orthogonal matrix containing the eigenvectors of \(\mathbf{X}^{\top}\mathbf{X}\), such that \(\mathbf{P}^{\top} = \mathbf{P}^{-1}\),
\item \(\mathbf{\Lambda} = \mathrm{diag}(\lambda_1, \lambda_2, \dots, \lambda_{2p})\) is a diagonal matrix containing the eigenvalues \(\lambda_1, \lambda_2, \dots, \lambda_{2p}\) of \(\mathbf{X}^{\top}\mathbf{X}\).
\end{itemize}

Using the eigen decomposition of \(\mathbf{X}^{\top}\mathbf{X}\), the covariance matrix \(\boldsymbol{\Sigma}_{\boldsymbol{\beta}}\) can be rewritten as:
\[
\boldsymbol{\Sigma}_{\boldsymbol{\beta}} = \left(\frac{1}{\sigma_{\epsilon}^2} \mathbf{X}^{\top}\mathbf{X} + \frac{1}{\sigma_{\beta}^2} \mathbf{I}_{2p} \right)^{-1}.
\]

Substituting \(\mathbf{X}^{\top}\mathbf{X} = \mathbf{P} \mathbf{\Lambda} \mathbf{P}^{\top}\), we have:
\[
\boldsymbol{\Sigma}_{\boldsymbol{\beta}} = \left(\frac{1}{\sigma_{\epsilon}^2} \mathbf{P} \mathbf{\Lambda} \mathbf{P}^{\top} + \frac{1}{\sigma_{\beta}^2} \mathbf{I}_{2p} \right)^{-1}.
\]

Since \(\mathbf{P}\) is an orthogonal matrix, the expression simplifies to:
\[
\boldsymbol{\Sigma}_{\boldsymbol{\beta}} = \mathbf{P} \left(\frac{1}{\sigma_{\epsilon}^2} \mathbf{\Lambda} + \frac{1}{\sigma_{\beta}^2} \mathbf{I}_{2p} \right)^{-1} \mathbf{P}^{\top}.
\]

The diagonal structure of \(\frac{1}{\sigma_{\epsilon}^2} \mathbf{\Lambda} + \frac{1}{\sigma_{\beta}^2} \mathbf{I}_{2p}\) allows us to compute its inverse as:
\[
\left(\frac{1}{\sigma_{\epsilon}^2} \mathbf{\Lambda} + \frac{1}{\sigma_{\beta}^2} \mathbf{I}_{2p}\right)^{-1} = \mathrm{diag}\left(\frac{1}{\frac{\lambda_1}{\sigma_{\epsilon}^2} + \frac{1}{\sigma_{\beta}^2}}, \frac{1}{\frac{\lambda_2}{\sigma_{\epsilon}^2} + \frac{1}{\sigma_{\beta}^2}}, \dots, \frac{1}{\frac{\lambda_{2p}}{\sigma_{\epsilon}^2} + \frac{1}{\sigma_{\beta}^2}}\right).
\]

Finally, the covariance matrix \(\boldsymbol{\Sigma}_{\boldsymbol{\beta}}\) is given by:
\[
\boldsymbol{\Sigma}_{\boldsymbol{\beta}} = \mathbf{P} \, \mathrm{diag}\left(\frac{1}{\frac{\lambda_1}{\sigma_{\epsilon}^2} + \frac{1}{\sigma_{\beta}^2}}, \frac{1}{\frac{\lambda_2}{\sigma_{\epsilon}^2} + \frac{1}{\sigma_{\beta}^2}}, \dots, \frac{1}{\frac{\lambda_{2p}}{\sigma_{\epsilon}^2} + \frac{1}{\sigma_{\beta}^2}}\right) \mathbf{P}^{\top}.
\]

To derive the full conditional distribution for $\sigma_{\beta}^2$, we will focus on the terms in the joint posterior density involving $\sigma_{\beta}^2$ as follows:

$$
\begin{aligned}
f\left(\sigma_{\beta}^2|\cdots\right)  & \propto \left(\sigma_{\beta}^2\right)^{-a_\beta-1} \exp \left(-\frac{b_\beta}{\sigma_{\beta}^2}\right) \times \left|\sigma_\beta^2\cdot\mathbf{I}_{2p}\right|^{-1 / 2} \exp \left\{-\frac{1}{2\cdot \sigma_{\beta}^2} \boldsymbol{\beta}^{\top} \boldsymbol{\beta}\right\}  \\
& = \left({\sigma_{\beta}^2}\right)^{-p} \left(\sigma_\beta^2\right)^{-a_\beta-1} \times \exp \left(-\frac{1}{2\cdot\sigma_{\beta}^2}\boldsymbol{\beta}^{\top}\boldsymbol{\beta}-\frac{b_\beta}{\sigma_\beta^2}\right) \\
& = \left(\sigma_\beta^2\right)^{-\left(a_\beta+p\right)-1} \times \exp \left(-\frac{1}{2 \cdot \sigma_\beta^2}\boldsymbol{\beta}^{\top}\boldsymbol{\beta}-\frac{b_\beta}{\sigma_\beta^2}\right) \\
& = \left(\sigma_\beta^2\right)^{-\left(a_\beta+p\right)-1} \times \exp \left(-\frac{1}{\sigma_\beta^2}\left[\frac{1}{2}\boldsymbol{\beta}^{\top}\boldsymbol{\beta}+b_\beta\right]\right) \\
& = \left(\sigma_\beta^2\right)^{-a_{\beta}^*-1} \exp \left(-\frac{b_{\beta}^*}{\sigma_\beta^2}\right),
\end{aligned}
$$
where $a_{\beta}^*=a_{\beta}+p$,  $b_{\beta}^*=b_{\beta}+\frac{1}{2}\times\boldsymbol{\beta}^{\top}\boldsymbol{\beta}.$
\noindent It has the IG kernel as above. Therefore, the full conditional distribution of $\sigma_{\beta}^2$ is the IG as follows:
\begin{itemize}
\item $\left[\sigma_{\beta}^2 \mid \cdots\right] \sim \mathrm{IG}\left(a_{\beta}^*, b_{\beta}^*\right),$ where $a_{\beta}^*=a_{\beta}+p$, \hspace{1mm} $b_{\beta}^*=b_{\beta}+\frac{1}{2}\times\boldsymbol{\beta}^{\top}\boldsymbol{\beta}.$
\end{itemize}

To derive the full conditional distribution for $\boldsymbol{\eta}$, we will focus on the terms in the joint posterior density involving $\boldsymbol{\eta}$ and the case $\boldsymbol{\beta}$ as follows:

$$
\begin{aligned}
f(\boldsymbol{\eta}|\cdots)  & \propto \exp \left\{-\frac{1}{2\cdot \sigma_{\epsilon}^2}\left(\boldsymbol{Z}-(\mathbf{X} \boldsymbol{\beta}+\mathbf{G} \boldsymbol{\eta}+\boldsymbol{\xi})\right)^{\top}(\boldsymbol{Z}-(\mathbf{X} \boldsymbol{\beta}+\mathbf{G} \boldsymbol{\eta}+\boldsymbol{\xi}))\right\} \times \exp \left\{-\frac{1}{2} \boldsymbol{\eta}^{\top} {\mathbf{\Sigma}_{\boldsymbol{\eta}}}^{-1} \boldsymbol{\eta}\right\} \\
& = \exp\Bigg\{-\frac{1}{2}\Bigg[\boldsymbol{\eta}^{\top}\left(\frac{1}{\sigma_{\epsilon}^2}\mathbf{G}^{\top} \mathbf{G}+{\mathbf{\Sigma}_{\boldsymbol{\eta}}}^{-1}\right) \boldsymbol{\eta}-2 \boldsymbol{\eta}^{\top} \left(\frac{1}{\sigma_{\epsilon}^2}\mathbf{G}^{\top}(\boldsymbol{Z}-\mathbf{X} \boldsymbol{\beta}-\boldsymbol{\xi})\right) \\
& + \frac{1}{\sigma_{\epsilon}^2}\boldsymbol{Z}^{\top} \boldsymbol{Z}-\frac{2}{\sigma_{\epsilon}^2} \left(\boldsymbol{\beta}^{\top} \mathbf{X}^{\top} + \boldsymbol{\xi}^{\top}\right) \boldsymbol{Z}+\frac{1}{\sigma_{\epsilon}^2}\left(\boldsymbol{\beta}^{\top} \mathbf{X}^{\top}+\boldsymbol{\xi}^{\top}\right) \left(\mathbf{X} \boldsymbol{\beta} + \boldsymbol{\xi}\right)\Bigg]\Bigg\} \\
& \propto \exp\Bigg\{-\frac{1}{2}\Bigg[\boldsymbol{\eta}^{\top}\left(\frac{1}{\sigma_{\epsilon}^2}\mathbf{G}^{\top} \mathbf{G}+{\mathbf{\Sigma}_{\boldsymbol{\eta}}}^{-1}\right) \boldsymbol{\eta}-2 \boldsymbol{\eta}^{\top} \left(\frac{1}{\sigma_{\epsilon}^2}\mathbf{G}^{\top}(\boldsymbol{Z}-\mathbf{X} \boldsymbol{\beta}-\boldsymbol{\xi})\right)\Bigg]\Bigg\}. 
\end{aligned}
$$

\noindent It has the multivariate normal kernel as above. Therefore, the full conditional distribution of $\boldsymbol{\eta}$ is the multivariate normal distribution as follows:
\begin{itemize}
\item $\left[\boldsymbol{\eta} \mid \cdots \right] \sim \mathrm{N}_{r}\left(\boldsymbol{\mu}_{\boldsymbol{\eta}}, {\boldsymbol{\Sigma}_{\boldsymbol{\eta}}}^{*}\right)$, \\ \hspace{0.5mm} where ${\boldsymbol{\Sigma}_{\boldsymbol{\eta}}}^{*}=\left(\frac{1}{\sigma_{\epsilon}^2} \cdot \mathbf{G}^{\top}\mathbf{G}+{\boldsymbol{\Sigma}_{\boldsymbol{\eta}}}^{-1}\right)^{-1},$ \\
$\boldsymbol{\mu}_{\boldsymbol{\eta}}={\boldsymbol{\Sigma}_{\boldsymbol{\eta}}}^{*}\left(\frac{1}{\sigma_{\epsilon}^2}\cdot \mathbf{G}^{\top}\left(\boldsymbol{Z}-\mathbf{X}\boldsymbol{\beta}-\boldsymbol{\xi} \right) \right),$
\end{itemize}
where $\boldsymbol{\mu}_{\boldsymbol{\eta}}$ has a dimension $r$ $\times$ $1$, and $\boldsymbol{\Sigma}_{\boldsymbol{\eta}}$ is in the positive definite matrix space $r$ $\times$ $r$.

To derive the full conditional distribution for ${\mathbf{\Sigma}_{\boldsymbol{\eta}}}$, we will focus on the terms in the joint posterior density involving ${\mathbf{\Sigma}_{\boldsymbol{\eta}}}$ as follows:

$$
\begin{aligned}
f\left({{\mathbf{\Sigma}_{\boldsymbol{\eta}}}}|\cdots\right)  & \propto  \left|{{\mathbf{\Sigma}_{\boldsymbol{\eta}}}}\right|^{-\frac{1}{2}} \exp \left\{-\frac{1}{2} \boldsymbol{\eta}^{\top} {{\mathbf{\Sigma}_{\boldsymbol{\eta}}}}^{-1} \boldsymbol{\eta}\right\} \times \left|{\mathbf{\Sigma}_{\boldsymbol{\eta}}}\right|^{-\frac{m+r+1}{2}} \exp \left[-\frac{1}{2} \operatorname{tr}\left(\boldsymbol{\Psi} {\mathbf{\Sigma}_{\boldsymbol{\eta}}}^{-1}\right)\right] \\
& = \left|{{\mathbf{\Sigma}_{\boldsymbol{\eta}}}}\right|^{-\frac{1}{2}} \exp \left\{-\frac{1}{2} \operatorname{tr}\left(\boldsymbol{\eta}^{\top} {{\mathbf{\Sigma}_{\boldsymbol{\eta}}}}^{-1} \boldsymbol{\eta}\right)\right\} \times \left|{\mathbf{\Sigma}_{\boldsymbol{\eta}}}\right|^{-\frac{m+r+1}{2}} \exp \left[-\frac{1}{2} \operatorname{tr}\left(\boldsymbol{\Psi} {\mathbf{\Sigma}_{\boldsymbol{\eta}}}^{-1}\right)\right] \\
& = \left|{{\mathbf{\Sigma}_{\boldsymbol{\eta}}}}\right|^{-\frac{1}{2}} \exp \left\{-\frac{1}{2} \operatorname{tr}\left(\boldsymbol{\eta}\boldsymbol{\eta}^{\top} {{\mathbf{\Sigma}_{\boldsymbol{\eta}}}}^{-1}\right)\right\} \times \left|{\mathbf{\Sigma}_{\boldsymbol{\eta}}}\right|^{-\frac{m+r+1}{2}} \exp \left[-\frac{1}{2} \operatorname{tr}\left(\boldsymbol{\Psi} {\mathbf{\Sigma}_{\boldsymbol{\eta}}}^{-1}\right)\right] \\
& = \left|{{\mathbf{\Sigma}_{\boldsymbol{\eta}}}}\right|^{-\frac{((m+1)+r+1)}{2}} \times \exp \left[-\frac{1}{2} \operatorname{tr}\left(\left(\boldsymbol{\Psi}+\boldsymbol{\eta} \boldsymbol{\eta}^{\top}\right) {\mathbf{\Sigma}_{\boldsymbol{\eta}}}^{-1}\right)\right] \\
& = \left|{{\mathbf{\Sigma}_{\boldsymbol{\eta}}}}\right|^{-\frac{(m^*+r+1)}{2}} \times \exp \left[-\frac{1}{2} \operatorname{tr}\left(\left(\boldsymbol{\Psi}^*\right) {\mathbf{\Sigma}_{\boldsymbol{\eta}}}^{-1}\right)\right], 
\end{aligned}
$$

\noindent where $\boldsymbol{\Psi}^* =\boldsymbol{\Psi}+\boldsymbol{\eta} \boldsymbol{\eta}^{\top}, m^*  =m+1$. It has the Inverse Wishart kernel as above. Therefore, the full-conditional distribution of ${\mathbf{\Sigma}_{\boldsymbol{\eta}}}$ is the inverse Wishart distribution as follows:
\begin{itemize}
\item $\left[{\mathbf{\Sigma}_{\boldsymbol{\eta}}}|\cdots\right] \sim \mathrm{InvWish}_r(\mathbf{\Psi}^*, m^*),$
where $\boldsymbol{\Psi}^* =\boldsymbol{\Psi}+\boldsymbol{\eta} \boldsymbol{\eta}^{\top}, m^*  =m+1$,
\end{itemize}
where $\mathbf{\Psi}$ and $\mathbf{\Psi}^*$ both denote positive definite scale matrices (which can be thought of as a matrix of sums of squares of a multivariate normal distribution), $m^*$ is the parameter that denotes the degrees of freedom and $r$ indicates the dimensions of ${\mathbf{\Sigma}_{\boldsymbol{\eta}}}$.

To derive the full conditional distribution for $\sigma_{\epsilon}^2$, we will focus on the terms in the joint posterior density involving $\sigma_{\epsilon}^2$ as follows:

$$
\begin{aligned}
f\left(\sigma_{\epsilon}^2|\cdots\right)  & \propto  \left|\sigma_\epsilon^2 \cdot \mathbf{I}_n\right|^{-1 / 2} \times \exp \left\{-\frac{1}{2\cdot \sigma_{\epsilon}^2}(\boldsymbol{Z}-\mathbf{X} \boldsymbol{\beta}-\mathbf{G} \boldsymbol{\eta}-\boldsymbol{\xi})^{\top}(\boldsymbol{Z}-\mathbf{X} \boldsymbol{\beta}-\mathbf{G} \boldsymbol{\eta}-\boldsymbol{\xi})\right\} \\
& \times \left(\sigma_\epsilon^2\right)^{-a_\epsilon-1} \exp \left(-\frac{b_\epsilon}{\sigma_\epsilon^2}\right) \\
& = {\left(\sigma_\epsilon^2\right)}^{-\frac{n}{2}} \left(\sigma_\epsilon^2\right)^{-a_\epsilon-1} \times \exp \left(-\frac{1}{2\cdot\sigma_\epsilon^2}(\boldsymbol{Z}-\mathbf{X} \boldsymbol{\beta}-\mathbf{G} \boldsymbol{\eta}-\boldsymbol{\xi})^{\top}(\boldsymbol{Z}-\mathbf{X} \boldsymbol{\beta}-\mathbf{G} \boldsymbol{\eta}-\boldsymbol{\xi})-\frac{b_\epsilon}{\sigma_\epsilon^2}\right) \\
& = \left(\sigma_\epsilon^2\right)^{-\left(a_\epsilon+\frac{n}{2}\right)-1} \times \exp \left(-\frac{1}{2\cdot\sigma_\epsilon^2}(\boldsymbol{Z}-\mathbf{X} \boldsymbol{\beta}-\mathbf{G} \boldsymbol{\eta}-\boldsymbol{\xi})^{\top}(\boldsymbol{Z}-\mathbf{X} \boldsymbol{\beta}-\mathbf{G} \boldsymbol{\eta}-\boldsymbol{\xi})-\frac{b_\epsilon}{\sigma_\epsilon^2}\right) \\
& = \left(\sigma_\epsilon^2\right)^{-\left(a_\epsilon+\frac{n}{2}\right)-1} \times \exp \left(-\frac{1}{\sigma_\epsilon^2}\left[\frac{1}{2}(\boldsymbol{Z}-\mathbf{X} \boldsymbol{\beta}-\mathbf{G} \boldsymbol{\eta}-\boldsymbol{\xi})^{\top}(\boldsymbol{Z}-\mathbf{X} \boldsymbol{\beta}-\mathbf{G} \boldsymbol{\eta}-\boldsymbol{\xi})+b_\epsilon\right]\right) \\
& = \left(\sigma_\epsilon^2\right)^{-a_{\epsilon}^*-1} \exp \left(-\frac{b_{\epsilon}^*}{\sigma_\epsilon^2}\right),
\end{aligned}
$$
where $a_{\epsilon}^*=a_{\epsilon}+\frac{n}{2}$,  $b_{\epsilon}^*=b_{\epsilon}+\frac{1}{2}\times(\boldsymbol{Z}-\mathbf{X} \boldsymbol{\beta}-\mathbf{G} \boldsymbol{\eta}-\boldsymbol{\xi})^{\top}(\boldsymbol{Z}-\mathbf{X} \boldsymbol{\beta}-\mathbf{G} \boldsymbol{\eta}-\boldsymbol{\xi}).$
\noindent It has the IG kernel as above. Therefore, the full conditional distribution of $\sigma_{\epsilon}^2$ is the IG as follows:
\begin{itemize}
\item $\left[\sigma_{\epsilon}^2 \mid \cdots\right] \sim \mathrm{IG}\left(a_{\epsilon}^*, b_{\epsilon}^*\right),$ where $a_{\epsilon}^*=a_{\epsilon}+\frac{n}{2}$, \\ $b_{\epsilon}^*=b_{\epsilon}+\frac{1}{2}\times(\boldsymbol{Z}-\mathbf{X} \boldsymbol{\beta}-\mathbf{G} \boldsymbol{\eta}-\boldsymbol{\xi})^{\top}(\boldsymbol{Z}-\mathbf{X} \boldsymbol{\beta}-\mathbf{G} \boldsymbol{\eta}-\boldsymbol{\xi}).$
\end{itemize}

To derive the full conditional distribution for $\sigma_{\xi}^2$, we will focus on the terms in the joint posterior density involving $\sigma_{\xi}^2$ as follows:

$$
\begin{aligned}
f\left(\sigma_{\xi}^2|\cdots\right)  & \propto \left(\sigma_{\xi}^2\right)^{-a_\xi-1} \exp \left(-\frac{b_\xi}{\sigma_{\xi}^2}\right) \times \left|\sigma_\xi^2\cdot\mathbf{I}_n\right|^{-1 / 2} \exp \left\{-\frac{1}{2\cdot \sigma_{\xi}^2} \boldsymbol{\xi}^{\top} \boldsymbol{\xi}\right\}  \\
& = \left({\sigma_{\xi}^2}\right)^{-\frac{n}{2}} \left(\sigma_\xi^2\right)^{-a_\xi-1} \times \exp \left(-\frac{1}{2\cdot\sigma_{\xi}^2}\boldsymbol{\xi}^{\top}\boldsymbol{\xi}-\frac{b_\xi}{\sigma_\xi^2}\right) \\
& = \left(\sigma_\xi^2\right)^{-\left(a_\xi+\frac{n}{2}\right)-1} \times \exp \left(-\frac{1}{2 \cdot \sigma_\xi^2}\boldsymbol{\xi}^{\top}\boldsymbol{\xi}-\frac{b_\epsilon}{\sigma_\xi^2}\right) \\
& = \left(\sigma_\xi^2\right)^{-\left(a_\xi+\frac{n}{2}\right)-1} \times \exp \left(-\frac{1}{\sigma_\xi^2}\left[\frac{1}{2}\boldsymbol{\xi}^{\top}\boldsymbol{\xi}+b_\xi\right]\right) \\
& = \left(\sigma_\xi^2\right)^{-a_{\xi}^*-1} \exp \left(-\frac{b_{\xi}^*}{\sigma_\xi^2}\right),
\end{aligned}
$$
where $a_{\xi}^*=a_{\xi}+\frac{n}{2}$,  $b_{\xi}^*=b_{\xi}+\frac{1}{2}\times\boldsymbol{\xi}^{\top}\boldsymbol{\xi}.$
\noindent It has the IG kernel as above. Therefore, the full conditional distribution of $\sigma_{\xi}^2$ is the IG as follows:
\begin{itemize}
\item $\left[\sigma_{\xi}^2 \mid \cdots\right] \sim \mathrm{IG}\left(a_{\xi}^*, b_{\xi}^*\right),$ where $a_{\xi}^*=a_{\xi}+\frac{n}{2}$, \hspace{1mm} $b_{\xi}^*=b_{\xi}+\frac{1}{2}\times\boldsymbol{\xi}^{\top}\boldsymbol{\xi}.$
\end{itemize}

To derive the full conditional distribution for $\boldsymbol{\xi}$, we will focus on the terms in the joint posterior density involving $\boldsymbol{\xi}$ as follows:

$$
\begin{aligned}
f\left(\boldsymbol{\xi}|\cdots\right)  & = (2 \pi)^{-n / 2} \cdot \left|\sigma_\epsilon^2 \cdot \mathbf{I}_n\right|^{-1 / 2} \times \exp \left\{-\frac{1}{2\cdot \sigma_{\epsilon}^2}(\boldsymbol{Z}-\mathbf{X} \boldsymbol{\beta}-\mathbf{G} \boldsymbol{\eta}-\boldsymbol{\xi})^{\top}(\boldsymbol{Z}-\mathbf{X} \boldsymbol{\beta}-\mathbf{G} \boldsymbol{\eta}-\boldsymbol{\xi})\right\} \\
& \times(2 \pi)^{-n / 2} \cdot \left|\sigma_\xi^2\cdot\mathbf{I}_n\right|^{-1 / 2} \times \exp \left\{-\frac{1}{2\cdot \sigma_{\xi}^2} \boldsymbol{\xi}^{\top} \boldsymbol{\xi}\right\} \\
&\propto \exp \left\{-\frac{1}{2\cdot \sigma_{\epsilon}^2}\left(\boldsymbol{Z}-(\mathbf{X} \boldsymbol{\beta}+\mathbf{G} \boldsymbol{\eta}+\boldsymbol{\xi})\right)^\top(\boldsymbol{Z}-(\mathbf{X} \boldsymbol{\beta}+\mathbf{G} \boldsymbol{\eta}+\boldsymbol{\xi}))  -\frac{1}{2\cdot \sigma_{\xi}^2} \boldsymbol{\xi}^\top \boldsymbol{\xi}\right\} \\
& = \exp\Bigg\{-\frac{1}{2}\Bigg[\boldsymbol{\xi}^{\top}\left(\frac{1}{\sigma_{\epsilon}^2} \mathbf{I}_n +\frac{1}{\sigma_{\xi}^2}\mathbf{I}_n\right) \boldsymbol{\xi}-2 \boldsymbol{\xi}^{\top} \left(\frac{1}{\sigma_{\epsilon}^2} \mathbf{I}_n\left(\boldsymbol{Z}-\mathbf{X} \boldsymbol{\beta}-\mathbf{G}\boldsymbol{\eta}\right)\right) \\
& + \frac{1}{\sigma_{\epsilon}^2}\boldsymbol{Z}^{\top} \boldsymbol{Z}-\frac{2}{\sigma_{\epsilon}^2} \left(\boldsymbol{\beta}^{\top} \mathbf{X}^{\top} + \boldsymbol{\eta}^{\top}\mathbf{G}^{\top}\right)\boldsymbol{Z}+\frac{1}{\sigma_{\epsilon}^2}\left(\boldsymbol{\beta}^{\top} \mathbf{X}^{\top}+\boldsymbol{\eta}^{\top}\mathbf{G}^{\top}\right) \left(\mathbf{X} \boldsymbol{\beta} + \mathbf{G} \boldsymbol{\eta}\right)\Bigg]\Bigg\} \\
& \propto \exp\Bigg\{-\frac{1}{2}\Bigg[\boldsymbol{\xi}^{\top}\left(\frac{1}{\sigma_{\epsilon}^2} \mathbf{I}_n +\frac{1}{\sigma_{\xi}^2}\mathbf{I}_n\right) \boldsymbol{\xi}-2 \boldsymbol{\xi}^{\top} \left(\frac{1}{\sigma_{\epsilon}^2} \mathbf{I}_n\left(\boldsymbol{Z}-\mathbf{X} \boldsymbol{\beta}-\mathbf{G}\boldsymbol{\eta}\right)\right)\Bigg]\Bigg\} \\
\end{aligned}
$$

It has the multivariate normal kernel as above. Therefore, the full conditional distribution of $\boldsymbol{\xi}$ is the multivariate normal distribution as follows:
\begin{itemize}
\item $\left[\boldsymbol{\xi} \mid \cdots \right] \sim \mathrm{N}_{n}\left(\boldsymbol{\mu}_{\boldsymbol{\xi}}, {\boldsymbol{\Sigma}_{\boldsymbol{\xi}}}\right)$, \\ \hspace{0.5mm} where ${\boldsymbol{\Sigma}_{\boldsymbol{\xi}}}=\left( \frac{1}{\sigma_\epsilon^2} {\mathbf{I}_n} +\frac{1}{\sigma_\xi^2} {\mathbf{I}_n}\right)^{-1},$ \\
$\boldsymbol{\mu}_{\boldsymbol{\xi}}={\boldsymbol{\Sigma}_{\boldsymbol{\xi}}}\left(\frac{1}{\sigma_\epsilon^2} {\mathbf{I}_n} \left(\boldsymbol{Z}-\mathbf{X} \boldsymbol{\beta}-\mathbf{G}\boldsymbol{\eta}\right)\right),$
\end{itemize}
where $\boldsymbol{\mu}_{\boldsymbol{\xi}}$ has a dimension $n$ $\times$ $1$, and $\boldsymbol{\Sigma}_{\boldsymbol{\xi}}$ is in the positive definite matrix space $n$ $\times$ $n$. The covariance matrix is:
\[
\boldsymbol{\Sigma}_{\boldsymbol{\xi}} = \left(\frac{1}{\sigma_{\epsilon}^2} \mathbf{I}_n + \frac{1}{\sigma_{\xi}^2} \mathbf{I}_n \right)^{-1}.
\]

Since \(\mathbf{I}_n\) is the identity matrix, the sum of scaled identity matrices is:
\[
\frac{1}{\sigma_{\epsilon}^2} \mathbf{I}_n + \frac{1}{\sigma_{\xi}^2} \mathbf{I}_n = \left(\frac{1}{\sigma_{\epsilon}^2} + \frac{1}{\sigma_{\xi}^2}\right) \mathbf{I}_n.
\]

Taking the inverse yields:
\[
\boldsymbol{\Sigma}_{\boldsymbol{\xi}} = \left(\frac{1}{\sigma_{\epsilon}^2} + \frac{1}{\sigma_{\xi}^2}\right)^{-1} \mathbf{I}_n.
\]

Simplifying further:
\[
\boldsymbol{\Sigma}_{\boldsymbol{\xi}} = \frac{\sigma_{\epsilon}^2 \sigma_{\xi}^2}{\sigma_{\epsilon}^2 + \sigma_{\xi}^2} \mathbf{I}_n.
\]

The mean vector is:
\[
\boldsymbol{\mu}_{\boldsymbol{\xi}} = \boldsymbol{\Sigma}_{\boldsymbol{\xi}} \left(\frac{1}{\sigma_{\epsilon}^2} \mathbf{I}_n \left(\boldsymbol{Z} - \mathbf{X} \boldsymbol{\beta} - \mathbf{G} \boldsymbol{\eta}\right)\right).
\]

Substituting \(\boldsymbol{\Sigma}_{\boldsymbol{\xi}} = \frac{\sigma_{\epsilon}^2 \sigma_{\xi}^2}{\sigma_{\epsilon}^2 + \sigma_{\xi}^2} \mathbf{I}_n\), we get:
\[
\boldsymbol{\mu}_{\boldsymbol{\xi}} = \frac{\sigma_{\epsilon}^2 \sigma_{\xi}^2}{\sigma_{\epsilon}^2 + \sigma_{\xi}^2} \mathbf{I}_n \cdot \frac{1}{\sigma_{\epsilon}^2} \mathbf{I}_n \left(\boldsymbol{Z} - \mathbf{X} \boldsymbol{\beta} - \mathbf{G} \boldsymbol{\eta}\right).
\]

Simplifying further:
\[
\boldsymbol{\mu}_{\boldsymbol{\xi}} = \frac{\sigma_{\xi}^2}{\sigma_{\epsilon}^2 + \sigma_{\xi}^2} \left(\boldsymbol{Z} - \mathbf{X} \boldsymbol{\beta} - \mathbf{G} \boldsymbol{\eta}\right).
\]

The full conditional distribution of \(\boldsymbol{\xi}\) is:
\[
\left[\boldsymbol{\xi} \mid \cdots \right] \sim \mathrm{N}_n\left(\boldsymbol{\mu}_{\boldsymbol{\xi}}, \boldsymbol{\Sigma}_{\boldsymbol{\xi}}\right),
\]
where:
\[
\boldsymbol{\Sigma}_{\boldsymbol{\xi}} = \frac{\sigma_{\epsilon}^2 \sigma_{\xi}^2}{\sigma_{\epsilon}^2 + \sigma_{\xi}^2} \mathbf{I}_n,
\]
\[
\boldsymbol{\mu}_{\boldsymbol{\xi}} = \frac{\sigma_{\xi}^2}{\sigma_{\epsilon}^2 + \sigma_{\xi}^2} \left(\boldsymbol{Z} - \mathbf{X} \boldsymbol{\beta} - \mathbf{G} \boldsymbol{\eta}\right).
\]

\section{Extension of Average Treatment Effect to the Continuous Treatment Setting}

{
	\hspace{5mm}The development above is written for a binary treatment, but the
	same mixture logic extends to a continuous exposure. For
	$a\in\mathcal A\subset\mathbb R$, let
	$\tilde Z_i(a,\tilde{\mathbf a}_{-i})$ denote the potential outcome under
	exposure level $a$, and define the dose--response function by
	\[
	\mu(a):=\mathbb E\{\tilde Z_i(a,\tilde{\mathbf a}_{-i})\}.
	\]
	For two exposure levels $a_1$ and $a_0$, define
	\[
	\tau(a_1,a_0):=\mu(a_1)-\mu(a_0).
	\]
	In Component 1, suppose
	\[
	g\{\mathbb{E}(Z_i\mid A_i,\mathbf x_i,\mathbf W_i,
	\boldsymbol\beta,\boldsymbol\eta)\}
	=
	\beta_0+\beta_A A_i+\mathbf{x}_i^\top\boldsymbol\beta_x+
	\mathbf W_i^\top\boldsymbol\eta.
	\]
	\noindent with scalar, vectors, and matrices defined in the main text. For a linear exposure effect on the link-function scale, the observational
	component carries the contrast
	\[
	\tau_{\mathrm{obs}}(a_1,a_0)=(a_1-a_0)\beta_A.
	\]
	For the assumption-fail component, let $\widetilde{\beta}_A$ denote the
	corresponding component-specific slope and define
	\[
	\tau_{\mathrm{mix}}(a_1,a_0)
	=(a_1-a_0)\widetilde{\beta}_A.
	\]
	Thus, the continuous-treatment AMUA posterior mixes the observational
	posterior for $\beta_A$ with a working assumption-fail distribution for
	$\widetilde{\beta}_A$. The HPD diagnostic may be reported for the resulting
	mixed slope, or equivalently for the induced contrast between fixed exposure
	levels. This notation keeps the tilde on the assumption-fail parameter while
	retaining the conventional notation $\beta_A$ for the observational-model
	coefficient.
	}

\begingroup

\section{Prior and Model Specifications}
\label{sec:supp_random_priors}

{
\noindent Let $\lambda_i^{(m)}$ denote the county-specific mortality rate under
model variant $m$, and let $\mu_i^{(m)}=N_i\lambda_i^{(m)}$ denote the
corresponding expected death count. A common notation for the four rate-scale
linear predictors is
\[
\log\lambda_i^{(m)}
=
A_i\beta_A^{(m)}
+
\vect{x}_i^\top\vect{\gamma}^{(m)}
+
I_P(m)\,\vect b(\widehat e_i)^\top\vect{\theta}^{(m)}
+
U_i^{(m)}.
\]
Equivalently, on the expected-count scale,
\[
\log\mu_i^{(m)}
=
\log N_i
+
A_i\beta_A^{(m)}
+
\vect{x}_i^\top\vect{\gamma}^{(m)}
+
I_P(m)\,\vect b(\widehat e_i)^\top\vect{\theta}^{(m)}
+
U_i^{(m)}.
\]
}
Here \(I_P(m)=1\) for NS+P and S+P and is zero otherwise. For S and S+P,
\(\vect U^{(m)}\) has the dependent proper-CAR structure. For NS and NS+P,
the dependent spatial component is absent. We make use of the same priors considered in Reich et al. (2021) outlined in Table~\ref{tab:supp_priors}.

The zero-centered regression priors and proper-CAR hyperpriors follow the
structure used in the spatial causal analysis of
\citet{ReichYangGuan2021}. 
\endgroup

\section{Computation and Stan Diagnostics}\label{sec:supp_computation}

\hspace{5mm}The exposure stage was fitted in OpenBUGS. Posterior summaries from that stage
were used to construct
\[
\widehat e_i
=
\left\{
A_i-
\left(
\vect{x}_i^\top\widehat{\vect{\alpha}}
+
\widehat V_i
\right)
\right\}^2,
\]
followed by a \(K=5\) B-spline basis \(\vect b(\widehat e_i)\). The four
models NS, NS+P, S, and S+P were then fitted separately in
Stan. The AMUA calculations are based on stored
post-warmup draws of \(\beta_A\). As an example, we provide the
he detailed Stan diagnostic material for
the S+P model. The run used three independent chains, with warmup=500, and sampling=5000
per chain. Warmup draws were retained for trace inspection, but posterior
summaries used only the \(3\times5000=15000\) post-warmup draws.

\noindent The posterior center is positive, but the 95\% HPD interval includes zero.
The small \(\widehat R\), the effective sample size, and the available traces
support stable post-warmup computation for this parameter.

\noindent
 Visual checks of the trace plots are consistent with
Table~\ref{tab:supp_stan_beta_A_summary} that indicate that convergence appears reasonable.



\begingroup
\section{Numerical Details to Approximate the ISE}
\label{sec:supp_ise}

\hspace{5mm}The ISE post-processing uses only the stored post-warmup marginal posterior
draws of \(\beta_A\) from NS, NS+P, S, and S+P. Each set contains 15000 draws. For each candidate scale parameter, all component densities are evaluated on a common
equally spaced grid
\[
b_1<\cdots<b_J,\qquad J=30000.
\]
The endpoints are chosen from the range containing all posterior draws together with the
\(10^{-5}\) and \(1-10^{-5}\) quantiles of the normal assumption-failure model stated in the main text.For model \(m\), let the raw kernel-density estimate from its posterior draws be
\[
\widetilde f_m(b)
=
\frac{1}{15000}
\sum_{r=1}^{15000}
K\!\left(
\frac{b-\beta_{A,m}^{(r)}}{h_m}
\right).
\]
\noindent
where $\bm{\beta}_{A,m}^{(r)}$ is the $r$-th simulated value from the posterior distribution for $\beta_{A}$ from model $m$. This is computed using the Gaussian kernel to define $K$, \texttt{stats::density(adjust=1)} in R, with the
model-specific default bandwidth denoted $h_{m}$. The resulting ordinates are interpolated
onto the common grid with \texttt{stats::approx} in R. The integration grid is
wider than the display range used in density figures, so plotted tail
truncation does not truncate the ISE calculation.

Since the finite-grid and interpolation can effect the estimated area under the curve numerically, each raw numerical KDE is re-normalized. Specifically, for model \(m\),
\[
\widehat A_m
=
\sum_{j=1}^{J-1}
(b_{j+1}-b_j)
\frac{\widetilde f_m(b_j)+\widetilde f_m(b_{j+1})}{2},
\qquad
\widehat f_m(b_j)=\frac{\widetilde f_m(b_j)}{\widehat A_m}.
\]

\noindent
The density for the assumption-fail component is first evaluated directly on the common grid for $b$ with
\[
\widetilde f_{F,m}(b;c)=\phi\!\left\{b;0,c^{2}s_{m}^{2}\right\}.
\]
The same normalization used above (i.e., $\hat{f}=\tilde{f}/\hat{A}$) is then applied on the finite common grid to obtain
\(\widehat f_{F,m}(b;c)\). For a normalized baseline density \(\widehat f_m\), we first form
\[
\widetilde f_{\mathrm{AMUA}}(b;c,\pi)
=
\pi \widehat f_m(b)+(1-\pi)\widehat f_{F,m}(b;c).
\]
The resulting mixture is again re-normalized on the same grid to obtain
\(\widehat f_{\mathrm{AMUA}}(b;c,\pi)\). 
Let $
d_j(c,\pi)
=
\widehat f_{\mathrm{AMUA}}(b_j;c,\pi)-\widehat f_{F,m}(b_j).$ 
The ISE is approximated by
\[
\operatorname{ISE}(c,\pi)
\approx
\sum_{j=1}^{J-1}
(b_{j+1}-b_j)
\frac{
d_j^2(c,\pi)+d_{j+1}^2(c,\pi)
}{2}.
\]
{which is minimized to find approximate optimal values of $c$ and $\pi$.} For each multiplier $c$ in a grid, we compute $
\pi^\star(c)
=
\arg\min_{\pi\in\{0,.01,\ldots,1\}}
\operatorname{ISE}(c,\pi),$
and
$c^\star=\arg\min_c \operatorname{ISE}(c,\pi^{*}(c)).$ Then we set $\pi^\star=\pi^\star(c^\star)$.

\endgroup

\begin{figure}[H]
\centering
\suppgraphic[0.98\textwidth]{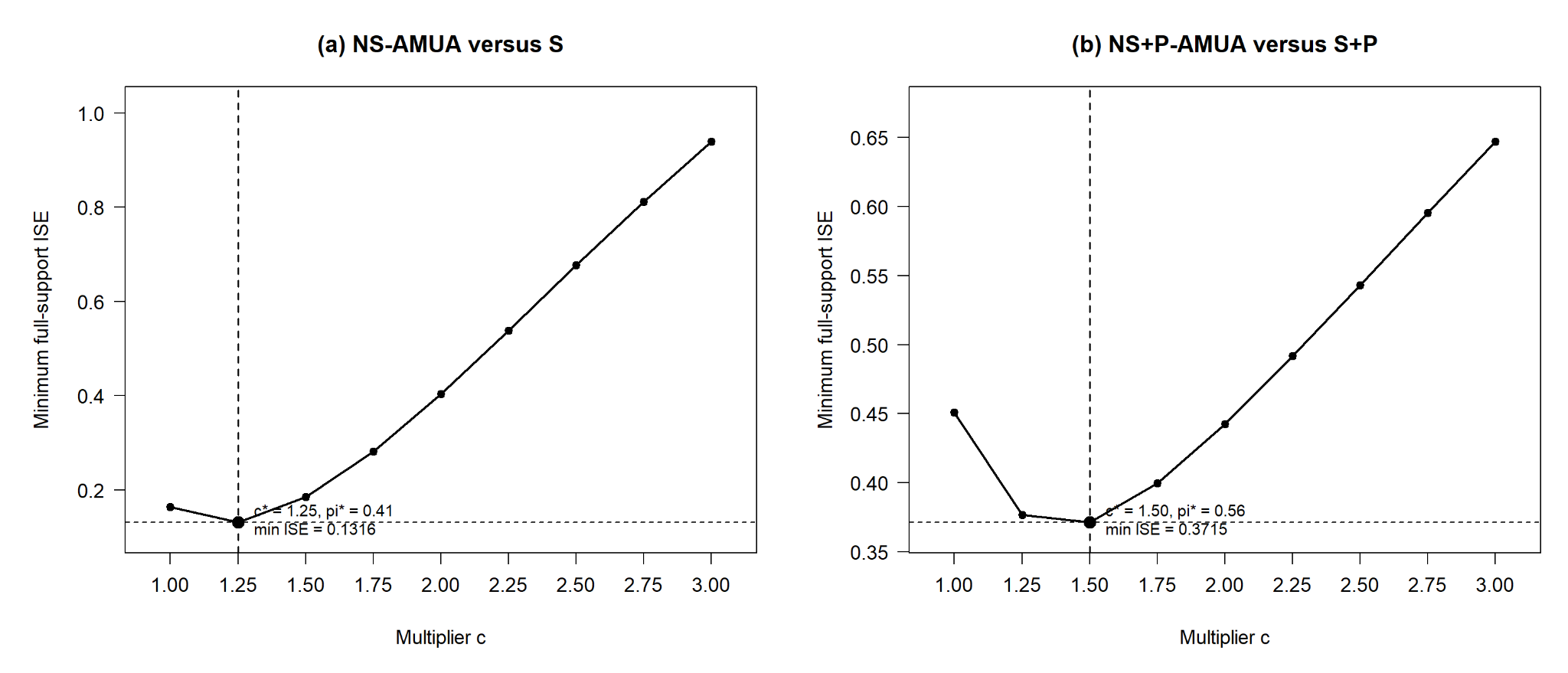}
\caption{Minimum  ISE after optimizing over \(\pi\) at each
failure-scale multiplier \(c\). The selected multipliers are \(1.25\) for
NS-AMUA versus S and \(1.50\) for NS+P-AMUA versus S+P.}
\label{fig:supp_min_ise_by_multiplier}
\end{figure}

\begin{figure}[H]
\centering
\suppgraphic[0.98\textwidth]{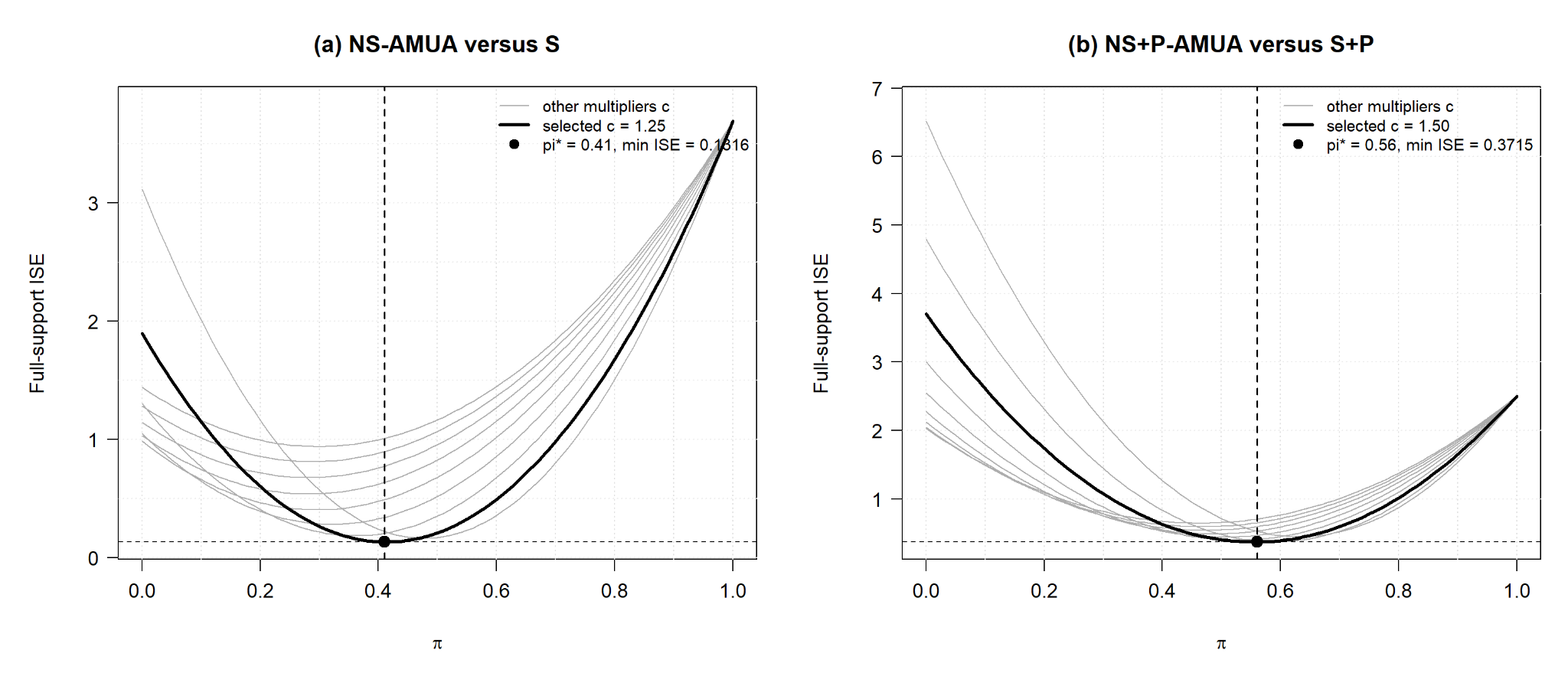}
\caption{Full-support ISE as a function of \(\pi\) for all candidate
multipliers. The emphasized curves correspond to the selected multipliers,
with minima at \(\pi^\star=0.41\) and \(0.56\).}
\label{fig:supp_ise_vs_pi}
\end{figure}

\begingroup
\section{Additonal $\pi^{*}$  Results}\label{sec:supp_calibration}

\hspace{5mm}For both NS-AMUA versus S and NS+P-AMUA versus S+P, the posterior mean, median, and standard deviation (Std) slightly. Plots of the ISE for various choices of $\pi$ and $c$ are given in Figures~\ref{fig:supp_min_ise_by_multiplier} and~\ref{fig:supp_ise_vs_pi}
These results show that the calibrated AMUA models (i.e., those based on $\pi^{*}$) reproduce much of the target center and spread,
although density shapes do not coincide exactly perfectly.

\hspace{5mm}These figures display the two nested selection steps. First, for each
candidate \(c\), ISE is minimized over \(\pi\). Second, the optimized minima
are compared across \(c\). The final selected mixtures are
\[
f^\star_{\mathrm{NS-AMUA}}(b)
=
0.41f_{\mathrm{NS}}(b)
+
0.59\phi(b;0,0.04033^2)
\]
and
\[
f^\star_{\mathrm{NS+P-AMUA}}(b)
=
0.56f_{\mathrm{NS+P}}(b)
+
0.44\phi(b;0,0.04953^2).
\]

\endgroup



\begin{table}[H]
\centering
\small
\caption{Prior distributions used in the PM$_{2.5}$--COVID-19 application.}
\label{tab:supp_priors}
\begin{tabular}{lll}
\toprule
Parameter & Prior distribution & Role\\
\midrule
\(\beta_A\)
& \(\mathcal N(0,10)\)
& PM$_{2.5}$ log-relative-risk coefficient\\
\(\vect{\alpha}\)
& \(\mathcal N(\vect 0,10\mat I)\)
& Exposure-model covariate coefficients\\
\(\vect{\gamma}\)
& \(\mathcal N(\vect 0,10\mat I)\)
& Outcome-model covariate coefficients\\
\(\vect{\theta}\)
& \(\mathcal N(\vect 0,10\mat I)\)
& GPS B-spline coefficients\\
\(\sigma_\epsilon^{-2}\)
& \(\mathrm{Gamma}(0.5,0.005)\)
& Exposure-model residual precision\\
\(\tau_V\)
& \(\mathrm{Gamma}(0.5,0.005)\)
& Exposure CAR precision\\
\(\tau_U\)
& \(\mathrm{Gamma}(0.5,0.005)\)
& Outcome CAR precision when applicable\\
\(\rho_V\)
& \(\mathrm{Uniform}(0,1)\)
& Exposure CAR dependence\\
\(\rho_U\)
& \(\mathrm{Uniform}(0,1)\)
& Outcome CAR dependence when applicable\\
\bottomrule
\end{tabular}
\end{table}

\begin{table}[H]
\centering
\caption{Stan posterior summary for \(\beta_A\) under the S+P
assumptions-holding model.}
\label{tab:supp_stan_beta_A_summary}
\resizebox{\textwidth}{!}{%
\begin{tabular}{lrrrrrrrrr}
\toprule
Quantity & Mean & SD & 2.5\% & Median & 97.5\% &
HPD lower & HPD upper & \(\widehat R\) & Bulk ESS\\
\midrule
\(\beta_A\)
& 0.0669 & 0.0553 & -0.0459 & 0.0679 & 0.1712
& -0.0455 & 0.1715 & 1.0017 & 748.8\\
\bottomrule
\end{tabular}%
}
\end{table}

\begin{table}[H]
\centering
\small
\caption{Selected AMUA calibrations under the
 ISE criterion.}
\label{tab:supp_selected_calibration}
\begin{tabular}{lrrrrr}
\toprule
Comparison & \(c^\star\) & $s_{m}$ & 
\(\pi^\star\) & Minimum ISE \\
\midrule
NS-AMUA vs.\ S
& 1.25 & 0.03226 & 0.41 & 0.1316 \\
NS+P-AMUA vs.\ S+P
& 1.50 & 0.03302 & 0.56 & 0.3715 \\
\bottomrule
\end{tabular}
\end{table}

\begin{table}[H]
\centering
\small
\caption{Posterior summaries for the NS, S, and the selected AMUA. HPD denotes the 95\% shortest interval.}
\label{tab:supp_selected_posterior_summary}
\resizebox{\textwidth}{!}{%
\begin{tabular}{llrrrrc}
\toprule
Comparison & Distribution & Mean & Median & Std & 95\% HPD &
Zero excluded?\\
\midrule
NS/S
& NS & 0.0762 & 0.0760 & 0.0323 & \([0.0130,0.1382]\) & Yes\\
& NS-AMUA & 0.0312 & 0.0320 & 0.0528 & \([-0.0666,0.1289]\) & No\\
& S & 0.0332 & 0.0341 & 0.0521 & \([-0.0688,0.1390]\) & No\\
\midrule
NS+P/S+P
& NS+P & 0.0992 & 0.0991 & 0.0330 & \([0.0386,0.1680]\) & Yes\\
& NS+P-AMUA & 0.0556 & 0.0679 & 0.0642 & \([-0.0722,0.1600]\) & No\\
& S+P & 0.0669 & 0.0679 & 0.0553 & \([-0.0454,0.1715]\) & No\\
\bottomrule
\end{tabular}%
}
\end{table}


\begin{figure}[H]
\centering
\suppgraphic[0.80\textwidth]{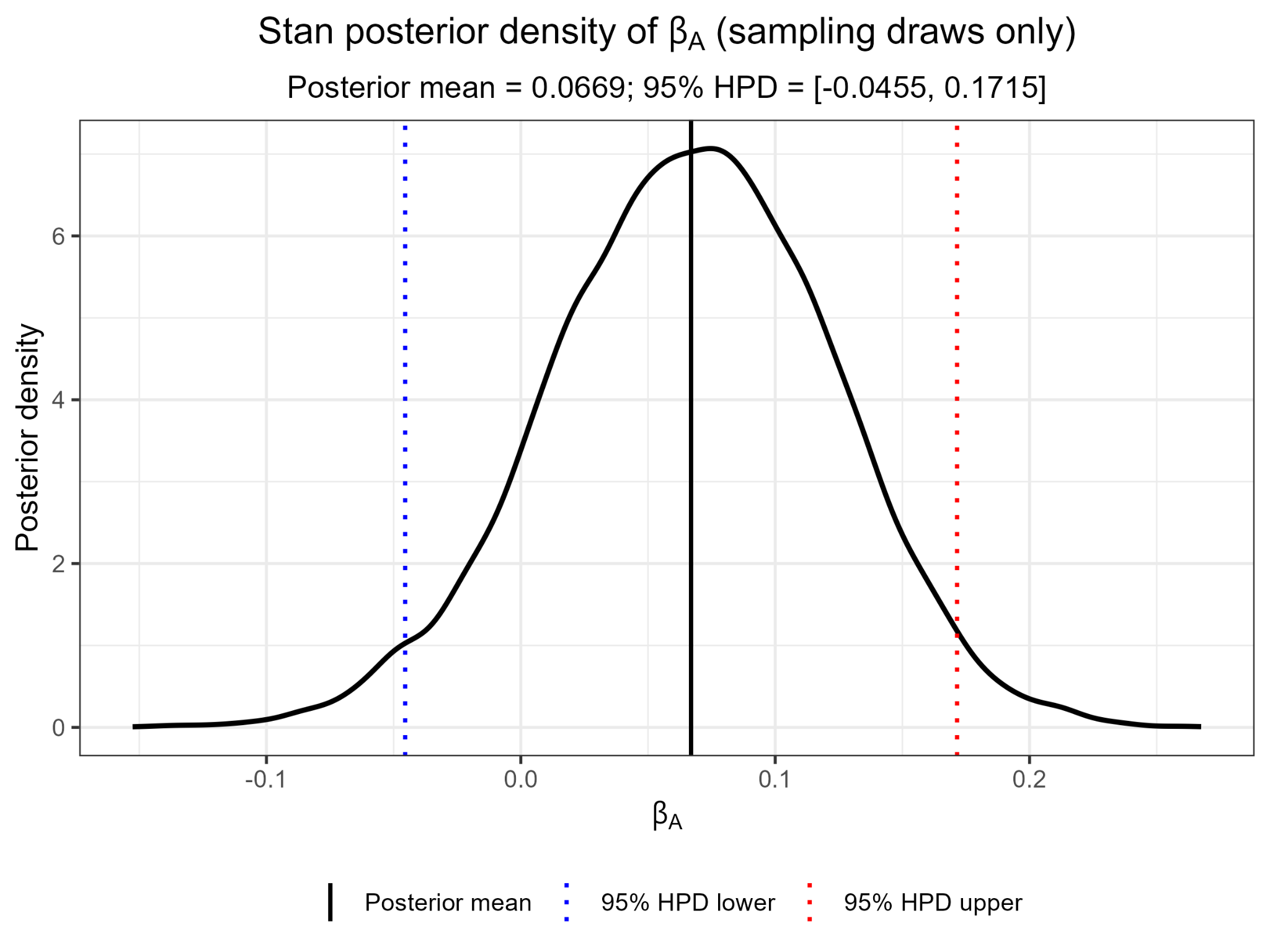}
\caption{Posterior density of \(\beta_A\) for the S+P model, based on pooled
post-warmup draws from the three Stan chains.}
\label{fig:supp_stan_beta_A_density}
\end{figure}

\begin{figure}[H]
\centering
\suppgraphic[0.70\textwidth]{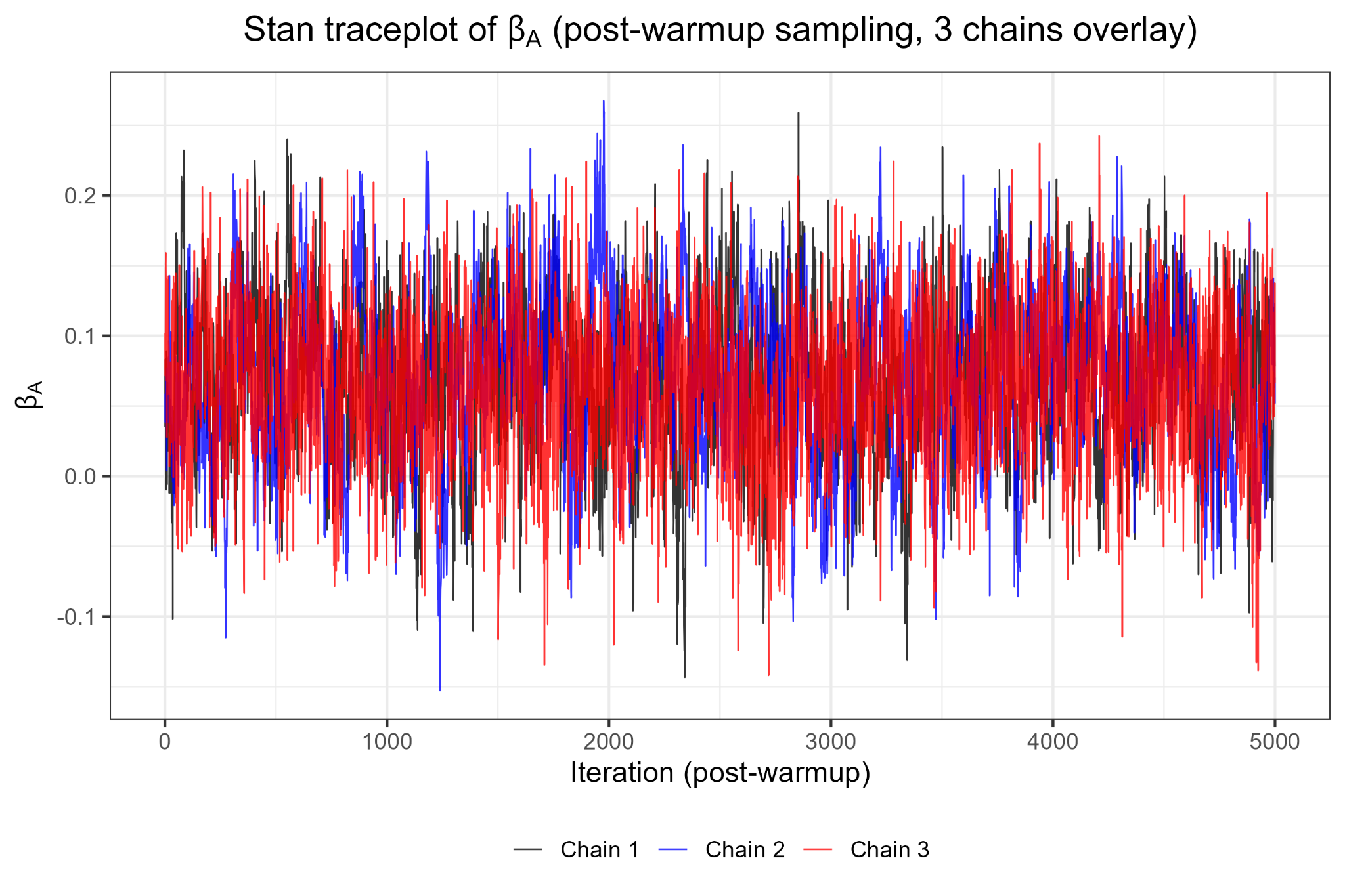}
\caption{Post-warmup Stan traceplot of \(\beta_A\) for the S+P model, with
the three chains overlaid.}
\label{fig:supp_stan_beta_A_trace_overlay}
\end{figure}

\begin{figure}[H]
\centering
\suppgraphic[0.70\textwidth]{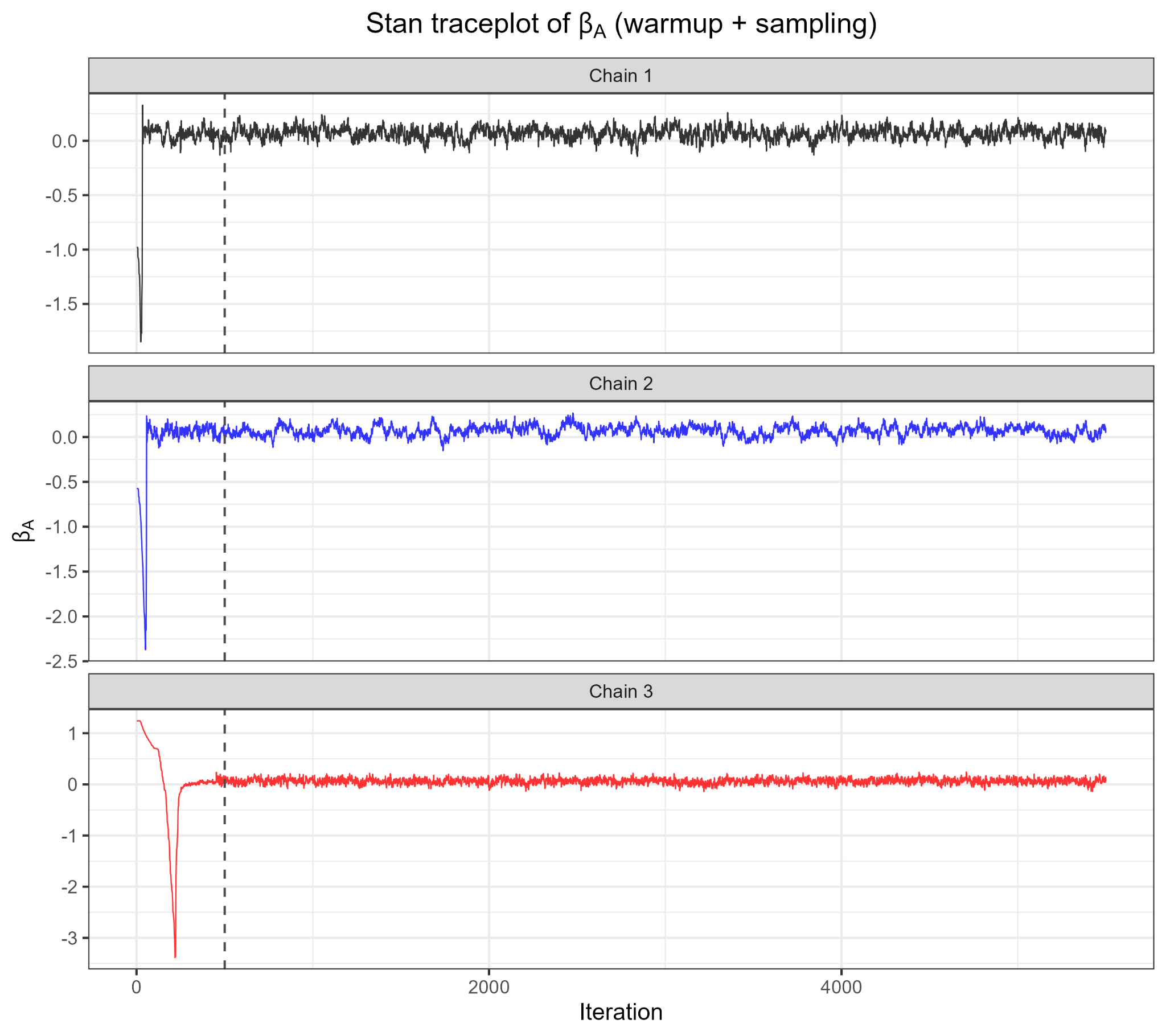}
\caption{Stan traceplots of \(\beta_A\) for the S+P model, including warmup
and post-warmup sampling periods.}
\label{fig:supp_stan_beta_A_trace_warmup}
\end{figure}

\end{document}